

\magnification\magstep  1
\parskip = 0.3 true cm
\baselineskip = 0.4 true cm
\overfullrule =0 pt
\hfuzz = 1 pt
\def \sqr#1#2{{\vcenter {\vbox{\hrule height.#2pt \hbox {\vrule width.#2pt
height#1pt \kern#1pt \vrule width.#2pt }\hrule height.#2pt}}}}
\def \square {\mathchoice \sqr34\sqr34\sqr{5.1}9\sqr{4.5}9}
\def \mow  { modular weights }
\def \fcnc { flavor changing neutral current }

\def \ew { electroweak }
\def \mssm {minimal supersymmetric standard model }
\def \IL { Ib\'a\~nez and L\"ust }
\def \BIM  {Brignole, Ib\'a\~nez and Mu\~noz }
\def \sugra {supergravity }
\def \wrt { with respect to }

\def\loe { low energy }
\def \km { Kac-Moody }
\def \susy { supersymmetry }
\def \susyq { supersymmetric }
\def \sig { $\sigma -$model }
\def \ws { world sheet }
\def\L {\Lambda }
\def\l {\lambda }
\def \t {\theta }
\def\a {\alpha }
\def\dh {\partial }
\def \d {\delta }
\def \D {\Delta }

\def \g {\gamma }
\def \G {\Gamma }
\def \O {\Omega }

\def \b {\beta }

\def \s {\sigma }
\def \e {\epsilon }
\def \cc { coupling constant }
\def \ccs {coupling constants }

\def \ren { renormalization }

\def \ud { {1 \over 2} }

\def \sm { standard model }

\def \eq {\eqno (  }

\centerline {\bf  Flavor changing neutral current constraints on
standard-like orbifold
models}
\vskip 1cm
\centerline {\it Philippe Brax and  Marc Chemtob }
\vskip 1cm

\centerline {\it Service de Physique Th\'eorique,
CE-Saclay, F-91191 Gif-sur-Yvette Cedex France
\footnote \ddag {\it Laboratoire de la Direction des Sciences
de la Mati\`ere du Commissariat \`a l'Energie Atomique }}

\vskip 1cm
{\bf Abstract }

\vskip 1cm

We examine for standard-like   orbifold
compactification models  the constraints due to
quarks and leptons  generation non-universality
of soft \susy
breaking interactions. We  follow the approach initiated by
\IL and developed  by \BIM .
This is based on a  locally \susyq \sig action  of
moduli and matter fields obeying
the  stringy duality symmetries. It is assumed
that the \loe fields of the \mssm  are in one-to-one correspondence
with string massless modes and that \susy breaking takes place simultaneously
with the lifting of flat directions for dilaton and  compactification
moduli fields.
The breaking of \susy is represented in terms
of dilaton and moduli  auxiliary field  components
 and, consistently with a
vanishing cosmological constant, is
parametrized in terms of  the dilaton-moduli
mixing angle $\theta $  and the
gravitino mass scale $m_g$.
The soft \susy breaking interactions
(gaugino masses, squarks and sleptons mass matrices,
scalars interactions  A and B coupling constants) are calculable as a
function of these parameters and of the  discrete set of
modular weight parameters
specifying  the modular transformation properties of the
\loe fields.
To establish the \fcnc constraints
we  solve  the \ren group
one-loop equations for the full set of gauge, Yukawa and \susy breaking
coupling constants.
A simplified version  is used in which one treats the contributions from
the third generation quarks and lepton Yukawa  couplings
exactly, while retaining
for the first and second generations
couplings
only the  leading order term in the large logarithm  variable. Numerical
results are obtained for the quantities
$ \Delta_{MN}= V_M  \tilde m^2_{MN} V_N^\dagger $,
corresponding
to the mass matrices of squarks and sleptons in the super-CKM basis,
for which  experimental bounds can be determined via
the super-box and super-penguin diagrams with gluino
or neutralino exchange.


\vfill\break
{\bf 1. Introduction }

The  string-based 4-d
\sugra $\s $-models inherit from
the   duality   symmetries of compactified strings  (see ref.[1] for a review)
an additional unbroken  symmetry corresponding to
discrete modular  groups which
act as  K\"ahler transformations
on the moduli and matter superfields [2].
The simplest case where
a general analysis of the
moduli space structure and  of the
modular  symmetry action can be made  is that
of  compactifications described by \ws $\sigma $-models
with (2,2) \susy [3].
No results of comparable rigor exist for  the
phenomenologically  more interesting
(0,2) models, with or without
Wilson lines. However, extensive analyses of these models exist
within the orbifold  toroidal
compactification framework  [4-8].  Here, string
tree-level correlators  are used to identify the K\"ahler potentials
of  the  string massless  moduli and matter modes
and to determine  modular
transformation laws of the  associated fields leading to assignment of
modular weights  [8]. String one-loop  correlators [9,10]
are also required  to
establish the cancellation of  \sig  duality
anomalies  which takes place as a result
of string mass  threshold effects
on gauge coupling constants  in combination with an analog of
the  Green-Schwarz  mechanism [11,12].

The restrictions  placed by duality  on \sugra $\sigma $-models
greatly enhance the predictive  power of studies
of \susy breaking within the so-called
standard-like models,   which are based on the  assumption
of a direct compactification  to the
minimal \susyq model.
One important aspect is to narrow the choice of  the
functional dependence on the compactification moduli
of the non-perturbative  component of
the superpotential which is supposed  to
lift flat directions in   dilaton and moduli  field space.
The main  hope, however,  is that these
non-vanishing VEVs (vacuum expectation values)
of scalar fields components of  dilaton and moduli are accompanied by
non-vanishing VEVs for the corresponding  auxiliary
fields components, thus  associating \susy
breaking with the same (as yet unknown)
mechanisms  that fix the gauge and Yukawa
\ccs of the
\loe theory [13-15].
The viability of this  picture has been examined in
several  recent works
within  the  hidden sector  gaugino condensation approach with a view to make
contact with  phenomenology [16,17].
 However, to avoid occurrence of
a  vanishing dilaton auxiliary field component,
$F_S=0$,  or a non-vanishing cosmological constant,
elaborate versions  must be used, in which
hidden matter  condensation or gaugino condensation from several
 gauge group  factors  play a role in the local \susy breaking
 [18]. Motivated by these limitations,
\BIM [19] have proposed   an alternative model independent
approach avoiding any
specific  assumption on the non-perturbative
superpotential except for the condition of non-vanishing
auxiliary fields components of
dilaton, moduli and possibly  matter  superfields, which
are parametrized in such a way as to  ensure  a vanishing
cosmological constant.
While  still rooted within   the gaugino condensation framework, this approach
is characterized by a  stronger emphasis on
auxiliary fields over  scalar fields, thus providing a transparent
representation of \susy breaking
in terms of mixing angles within the  goldstino and the
gravitino mass.  However,  the scalar  VEVs stand out
as extra free parameters in this approach.

The \loe  limit  of
standard-like models
can then be   derived on minimal grounds
by assuming a one-to-one correspondence
between  string massless modes and the  \loe
(quarks, leptons, Higgs)  particles  and, in the familiar way,
expanding the \sugra  \sig model action
in powers of the inverse Planck mass $1/M_P$.
 The \susy breaking parameters
(gaugino masses,  scalars masses,
bilinear and trilinear  scalars  interactions  couplings)
can be explicitly calculated
 in terms of  the  dilaton and moduli  scalar and auxiliary
 components  VEVs
 and of  the fields modular weights.

 One characteristic  feature of the predicted soft \susy breaking \ccs
is the suggested  presence of large  non-universal
contributions with respect to the
quarks and leptons generation quantum numbers [8].
Such generation dependence of  squarks and  sleptons mass matrices and
of A-matrices for couplings of squarks and sleptons to Higgs bosons
could  contribute to  \fcnc  processes
at an observable level, a fact which can be used under appropriate
assumptions to set upper bouds on  generation dependence [20,21].
However,  the relevant physical observables in  \sugra models are those
obtained after summing the large logarithms
contributed by  radiative corrections.
This requires solving  the \ren group flow equations down
to the \ew scale, using the soft parameters as
boundary conditions   at the
string unification scale [22,23].
While the generation independent interactions of gauginos
reduce  whichever large non-universality present  at the string scale,
Yukawa interactions have an opposite effect.
It is then mainly a quantitative  issue to determine what
residual non-universal dependence
is not wiped out  at the \ew scale.
Also,  the fact that the  quantities which
enter in the evaluation of
super-box or super-penguin diagrams
involve  the tranformation matrices which diagonalize
the quarks and leptons mass matrices,  implies that the relevant
observables are also sensitive
to the  flavor structure of Yukawa interactions.

The issue of
universality of  \susy breaking is closely tied with the flavor problem.
Conceptually, it is difficult to reconcile the important flavor
asymmetries present  in Yukawa interactions of quarks
and leptons with the postulate
of a  spontaneous \susy breaking  independent of flavor or generation
quantum numbers [24].
Non-universality also poses an acute naturalness problem  to \sugra models
[25,26]. This arises in  grand unified theories embedding  of the  \mssm
where radiative corrections from
decoupling  of superheavy modes can induce
contributions to mass matrices of scalar
superpartners in generation space   which are
misaligned with those arising from
Yukawa couplings of
\loe modes [20].
The early envisaged  possibility to limit the size of non-universal terms
on the basis of
flavor (horizontal) discrete or continuous
symmetries has been revived recently in a number
of interesting works  [27-30]. Relaxing the  universality  restriction
also plays an important role in widening the allowed parameter space of
the \mssm  [31].
Note however that string  theory   interpretation  of  non-universality
differs from that of field theory   in two respects: (i) Non-universality
is associated with generation dependent assignment of
modular weights to  \loe fields taking discrete rational values
rather than  with  the
spontaneous breakdown of an horizontal symmetry;   (ii) It
sets in at the larger string unification   scale which should
ease its  smearing
by flavor-blind   radiative  corrections.
Thus, string theory provides us with a
framework to accommodate large non-universality as well as
mechanisms to reduce their  observable effects.
Related recent  works on this  subject are refs.[32,33]

Our main goal in this work is to examine  the  implications
of the experimental  \fcnc bounds  [20,34]
concerning the presence of a generation  non-universality in the
dimension-two scalars mass matrices, $ \tilde m^2_{MN}, [M,N=L,R]$,
and the  dimension-three  scalars  couplings matrices $A^x_{ij}, [x=u,d,e]$.
In the super-CKM basis, with flavor diagonal
Yukawa couplings of quarks and squarks
or leptons and sleptons to gauginos,
the generation  dependence in the
super-box or super-penguin diagrams,
propagating  superpartners of \loe particles,
involve the quantities,
$ \Delta_{MN}= V_M  \tilde m^2_{MN} V_N^\dagger $, where $V_M, [M=L,R]$
are transformations which diagonalize the quarks and leptons mass matrices.
We shall restrict  consideration to
the standard-like orbifold compactification leading to the
\mssm .
The local \susy breaking parameters are then  determined as a  function
of the  dilaton and moduli auxiliary  and scalar fields components
at the  large string unification scale. To
compare with upper bounds
on the parameters $\D_{MN}$, we need to integrate the scale evolution
equations.
As is well known, the corrections from gauge bosons and fermions  loops,
being flavor-blind,   cause important dilution of flavor mixing  effects [26].
Thus, non-universality  should be strongly suppressed
in a  dilatonic goldstino vacuum, $\cos \t \approx 0$ [19].
A naive estimate [19]  including only gauge
interactions in the scale evolution
yields a large  upper bound on the cosine of the  mixing angle,
$\vert \cos \t \vert \le 1/\sqrt 3  $, corresponding to a  narrow range
of values for the  dilaton-moduli mixing angle,
$ 90^\circ \le  \t  \le 55^\circ $.
An independent source of flavor mixing  is also present  in the
trilinear scalar interactions, which are affected too
by  the \ren group evolution.

The contents of the paper are organized in 4  sections.
In Section 2,  we  recall  the main ingredients of
\sugra \sig effective actions incorporating
duality symmetries, as  developed by
\IL [8] ( Subsection 2.1),  and  of \susy breaking in this
framework, using the parametrization of  \BIM [19] (Subsection 2.2).
Next, we present
the approximate  version used for the one loop
\ren group equations  in which
the  contributions from  the third
generation particles (top and bottom quarks and $\tau$-lepton) are
included exactly, while
those of the first and second generations are included perturbatively
(Subsection 2.3). Finally we discuss the choice at string scale for the
Yukawa coupling matrices
(Subsection 2.4).
 In Section 3,  we present and discuss our numerical results.
 In Section 4, we summarize  our main conclusions and suggestions for
a future improved treatment of the problem.

{\bf 2.  Duality symmetry and \susy breaking interactions}

{\bf 2.1 \sig effective action }

By  moduli space(s) of a superstring theory  one  means the
configuration space  sector(s) of the
associated \ws  field theory spanned by marginal operators deformations
preserving the superconformal symmetry.  This  space  constitutes the
vacuum manifold of the
\loe 4-d  effective field theory.
The moduli space of string-based $\s $-models \sugra are parametrized by:
(1) the space-time  dilaton field, $S(x)$;
(2) the  compactification
moduli  fields, $ M(x), \bar M (x)$,  associated
to  deformations of the  K\"ahler    and complex structures of the
6-d internal space manifold;
(3)  the  matter-like background  fields,
$A^W_\g$, associated to Wilson  gauge
flux  lines winding around non-contractible loops in
the lattices of $E_8\times E_8
$ root space and  internal 6-d space-time and other possible massless
fields associated with duality [6].
Orbifolds (see refs.[35] for  reviews)
$\O = R^6/(\L \times P)$
represent special exact solutions
for  an internal space given by a  6-torus of lattice $\L $
with a discrete point symmetry group $P$
generated by  elements
of  the $SU(3)$ holonomy subgroup of the 6-d tangent space group
$SO(6)$. The  point groups of
abelian, $Z_N $ or $Z_N \times Z_M$,  orbifolds
are generated by elements  which can be represented in   a
complex basis as:
$ \t =diag( e^{2\pi i \t_i} ) , \quad [\sum_i \t_i=1] $
with  complex conjugate indices $i
,\bar i =1,2,3$ labeling
coordinates, $X^{i=1,\bar 1}={1\over \sqrt 2}X^{\mu =1\pm i2}, \cdots $,
and spinors.
The relevant compactification moduli fields  constitute then a
subset of the moduli matrices,
$[T_{i\bar j}, \bar T_{i\bar j }] , \quad
[U_{i\bar j}, \bar U_{i\bar j}],   \quad [i, j=1,2,3], $
invariant under $P$.
For convenience, we shall restrict ourselves
to the so-called generic orbifold case,
characterized as allowing only diagonal K\"ahler structure  moduli,
$T_i, \bar T_i [i=1,2,3] $, or a subset of these, while
excluding entirely complex structure moduli $ U_i, \bar U_i $  and matter
background fields $A^W_{\g }$. (See ref.[9] for  the list of
allowed moduli and ref.[8] for  the list of \sig  moduli spaces).
As in most  of the
recent studies, we also  shall specialize to
decomposable  6-d tori with a  metric tensor given by a
direct product of metric tensors  for
three orthogonal  2-d tori, or complex planes, labelled by indices $i, \bar i
=1,2,3$. (Calculations with
non-decomposable tori are considered in refs.[36,37]).

The  K\"ahler potential of the effective $N=1$ \sugra effective  action
can be derived in general form for the case  of \ws   with (2,2) \susy [3].
The part involving untwisted moduli is given in exact form as:
$$ K_{moduli}(S,\bar S,M,\bar M, A^W_\g ,A^{W \dagger }_\g  )
=-\sum_i \log \bigg (T_i+\bar T_i
- \sum_{\g }A^{W\dagger }_\g
 \prod_a e^{2g_a V_a(R_\g )} A^W_\g
\bigg ) $$$$-\sum_i \log(U_i+\bar U_i)
-\log (S+\bar S),\eq 1)$$
while the part involving  untwisted and twisted sectors matter
superfields, denoted generically by $A_\a= A_\a +\t \chi_\a +\t^2 F_\a $,
where the index $\a $ designates whatever quantum numbers (gauge, family,...)
are needed to characterize string modes,
is given in an expansion in powers of ${A_\a\over M_P}$ as:
$$ K_{matter}(M,\bar M,A_\a, A^\dagger _\a  )=\sum_\a \prod_i (T_i+\bar
T_i)^{n_\a^i}
\prod_m(U_m+\bar U_m)^{l_\a^m} \bigg (A^\dagger _\a \prod_a e^{2g_a V_a(R_\a )}
A_\a\bigg )+\cdots ,\eq 2) $$
where $n_\a^i, l_\a ^m$ are  fields modular weights, rational
numbers characterizing the modular group transformation laws.
For each matter field, we have inserted
its minimal gauge  coupling interaction, such that  each group
factor $G_a$, with coupling constant $g_a$, is associated
the real superfield $V_a(R_\a )= -iV_a^{a'} T^{a'}(R_\a )$,
with $T^a (R_\a )$, the Lie algebra generators in the representation $R_\a $
subject to the normalization convention
$ Tr(T_aT_{a'}
)=\ud c(R_\a ) \d_{a a'} ,$
 where $ c(R_\a)$  is the Dynkin index of representation $R_\a$.
Note that the absence of off-diagonal kinetic terms in eq.(2)
is an exact result
for (2,2) models which is expected to  remain true in general cases.
The above formulas are found to hold
for the  wider class of  \ws (0,2) models only
for orbifolds.  We  shall  restrict
ourselves to this case in the sequel.  The duality symmetries are
realized for the diagonal moduli, $T_{i\bar i}= T_i$,  by the
modular symmetry group
$\G = \big [SL(2,Z)\big ]^3$ (or an appropriate subgroup),
whose action  on the moduli and matter fields in the supergravity
basis (as opposed to the string vertex operators basis ) [6] is specified
by the transformations:
$$ S\to S, \quad  \quad T_i\to {a_iT_i-ib_i\over ic_iT_i+d_i},\quad
A_\a \to \prod_i
(ic_iT_i+d_i)^{n_\a^i} A_\a , \eq 3)$$
where $(a_i,b_i,c_i,d_i)\in Z, \quad  a_id_i-b_ic_i=1$, for $i=1,2,3$,
define elements of $SL(2,Z)_{T_i}$ and the modular weights are
given  as [8]:
$n_\a^i=-\d_{\a i} , $ (untwisted sector $\a =i=[1,2,3]$) and
$n_\a^i=-(1-\t^i+p^i-q^i)$,
(twisted sectors),
$p^i, q^i$ being  the  number of complex conjugate
left-moving sector oscillator excitations
of type $\a^i_{m-\t^i}, \tilde \a^i_{m+\t^i}$.
Note that the background fields $A_\g ^W$ transform as untwisted states
and that for complex structure moduli analogous formulas
hold for  the associated  modular weights $l_\a ^m$
with the interchange of $p^i $ with $  q^i$.

Each gauge symmetry group factor $ G_a$
of the full gauge group, $\prod_a G_a$,
is associated a holomorphic gauge function $f_{aa'}$, defined  in the adjoint
representation.
In the globally \susy limit,
with the field strength chiral
superfield defined as,  $-iW^a_\a  T^a =
{1\over 4} \bar D^2 e^{-V_a} D_\a e^{V_a}=
-{i\over 4}(i\l^a_\a +\cdots ) T^a$,
this contributes   the familiar  Lagrangian:
 $$ L_{gauge}= {1\over 4} \sum_a
\int d^2\t W^{a\a }W_\a ^{a'} f_{aa'}(S,T_i) +c.c., \eq 4)  $$
such that
$f_{aa'} (S,M_i)=k_a S\d_{aa'}, $ with  the dilaton VEV being related to the
string coupling constant as
$<S>=1/g_X^2$
 and $k_a$  being the level of the \ws  \km algebra for the group $G_a$.
The only requirement from duality  on the superpotential
is that it be a modular form of $SL(2,Z)_{T_i}$
of weight $-1$, corresponding to
the  modular
group acting  as  the K\"ahler transformation,
$K\to K+ f_i+\bar f_i
, W\to e^{-f_i }  W ,\quad  [f_i= \log (ic_i T_i +d_i), \quad \bar f_i=f^\star
_i] $.

The consideration of  string and \sig  loop expansions  provides
important information on the moduli dependence of
the effective
action. Realization of duality as  a K\"ahler
transformation,
 entails   the presence of   chiral $U(1)_{T_i}$ R-symmetries
transforming  matter and   gauge fermions as,
$\psi \to \psi e^{-iw_i(\psi )
Im f_i } ,$  with weights $  w_i(\l^a)={1\over 2} , \quad
w_i(\chi_\a )=-{1\over 2} (1+2n^i_\a )$.
For the diagonal modular subgroup, $ f=\sum_i f_i, w(\l_a)={1\over 2},
w(\chi_\a ) =-{1 \over 2} (1+{2\over 3} n_\a ), \quad n_a =\sum_i n_\a ^i $.
 At the quantum level these symmetries
are affected by \sig triangle anomalies
involving non-local couplings of gauge bosons  to
K\"ahler and matter   composite connections,
$$\eqalign {L_{nl}=&-{1\over 4}\sum_a \int d^2\t W^aW^a{1\over (4\pi )^2}
{\bar D^2 D^2\over 16 \square }\bigg [\ud \bigg (c(G_a)-\sum_{R_\a }
c(R_\a )\bigg ) K_{mod} \cr
&+\sum_{R_\a } c(R_\a )\log det_{\a \bar \b }
(K_{mat})_{\a \bar \b } \bigg ] +c.c. \cr } $$
(Analogous \sig gravitational anomalies are also present.)
The duality anomalies contain a
gauge group independent (universal) part which can be cancelled by an
analog of the  Green-Schwarz
mechanism [11,12].
This is implemented  in the dilaton  chiral  multiplet formulation by
changing the dilaton  field variable  to [8,12]:
$  S\to S^{new}= S-{1\over (4\pi )^2} \sum_i\d^i_{GS}\log (T_i+\bar T_i)$,
such that  the new field transforms under modular transformations
as: $S^{new}\to S^{new}+{2\over (4\pi )^2} \sum_i \d_{GS}^i (f_i-\bar f_i)$,
while $S$ is inert.
Performing the implied substitution  in eq(1), and
suppressing the subscript 'new',  the
corrected   dilaton  K\"ahler
potential  becomes: $$\log (S+\bar S)\to  \log (Y)\equiv
\log \bigg (S+\bar S+{2
\over (4\pi)^2} \sum_i\d_{GS}^i\log (T_i+\bar T_i) \bigg ), \eq 5) $$
where the modular invariant function $Y$ is
interpreted as an effective dilaton field whose VEV defines
a renormalized string
coupling constant field, $<Y>=2/g^2_X$.
At string one-loop level, there also arises mass threshold corrections in the
gauge \ccs which, after changing the dilaton field
variable in the tree level term, leads to corrected gauge  functions [9]:
$$f_{aa'} (S,T_i)\equiv f_a(S,T_i)\d_{aa'}=\d_{aa'} \bigg [ k_aS+\sum_i{\tilde
b_a^{'i} \over
(4\pi)^2}\log  \eta^4(T_i)
\bigg ], \eq 6)$$ where
$\eta (T) $ is Dedekind function (automorphic modular form of weight $\ud $)
and $\tilde b_a^{'i}= b^{'i}_a-k_a\d_{GS}^i,  $ with
$b^{'i}_a=\ud [c(G_a)-\sum_\a (1+2n_\a ^i) c(R_\a )] $,
corresponding to the  $\b $-function slope parameters for the
N=2 suborbifolds associated with the subgroup of the point symmetry
group leaving the $i-$plane fixed.
Using eqs.(1) and (2)  to write the above non-local effective  Lagrangian
for the \sig modular anomalies as:
$$L_{nl}=-{1\over 4(4\pi )^2}\sum_a \int d^2\t W^aW^a
{\bar D^2 D^2\over 16 \square }
\sum_i b^{'i}_a \log (T_i+\bar T_i) +c.c. ,  $$
we see that the modular transform of $L_{nl}$ is composed of a
moduli-dependent term  which is cancelled  by threshold
corrections (second  term in the gauge
function,  eq.(6)) and a universal   moduli-independent term
which is cancelled by virtue of the Green-Schwarz   dilaton-moduli
mixing (first tree
level term in eq.(6)).
Parenthetically, we note that  an analogous cancellation
also takes place for the QCD triangle anomalies affecting
the  K\"ahler R-symmetries   $U(1)_{T_i}$ which are compensated
by the breaking term occurring in   threshold corrections,
$(\t_a)_{thres} =-(4\pi )^2 Im f_a (S, T_i)=
-b^{'i}_a \log \bigg ({\eta (T_i)
\over \bar \eta (T_i)}\bigg )^2 $. This term arises [37] because any modular
invariant coupling term  quadratic in colored fields can be
transfered  via a chiral rotation into
a contribution to the $\t $-vacuum parameter,
whose variation under $U(1)_{T_i}$, $\d_i \t_a = \ud
\bigg ( c(G_a)-\sum_\a
(1+2n^i_\a ) c(R_\a )\bigg ) (f_i-\bar f_i)$, is exactly canceled by that of
threshold corrections, $ \d_i (\t_a)_{thres}= - b^{'i}_a (f_i-\bar f_i)$.

To simplify the  discussion  of moduli dependence, it is
convenient to limit the number of
unknowns by  restricting  to a one-dimensional
direction in moduli  space, corresponding to the  overall volume
modulus field $ T(x)$.
For an isotropic orbifold with three equivalent 2-tori,  the overall modulus
is defined by the restriction:
 $T_1(x)=T_2(x)=T_3(x)=T(x)$, and
transforms under the diagonal subgroup $SL(2,Z)$  of
$\big [SL(2,Z)\big ]^3$,  whose \mow are given by the sums,
$n_\a =\sum_in_a^i.$ The overall modulus for an
anisotropic orbifold can be defined analogously by the condition:
$r_1T_1=r_2T_2=r_3T_3=T$, where  $r_i$ are   real
parameters   characterizing the geometrical shape of the 6d-torus which
obey  $\prod_i r_i=1$, by  virtue of volume conservation.
The squeezed orbifold with $T_1 \gg T_2,T_3 $ has $ r_1\ll r_2,r_3$.
The diagonal moduli VEVs are
related to the  radii of the 2-d tori
as: $ Re (T_i)=c_NR_i^2/( (4\pi )^2)\a' ) $,
with $\a'$ the string slope parameter and
$c^2_N \propto  det (G) ,$ with  $  G$ the  6-torus metric tensor.
The $c_N $ are calculable  constants  depending on the
$Z_N $ orbifold order. For instance,
$ c_3=c_6= \sqrt 3 , \quad c_4= 2\sqrt 2 $.
(Applications dealing with non-decomposable 6-d tori
are developed in refs.[36].)  Naturally,
the values at which  these  radii,
or the corresponding  overall modulus and  anisotropies,  settle is
determined by  compactification dynamics which
lies outside the  present framework. However, because compactification
means a choice of vacuum,
minimization of the \sugra scalar potential will
place certain  conditions on these VEVs.
Thus, for decomposable tori,  the equivalence of different planes imposes
a discrete permutational symmetry among the allowed moduli.
An anisotropic solution could  then arise in either of the following two
situations:
(1) A subset of the  diagonal moduli is allowed.
The $Z_4$ orbifold, for example,  allows for two diagonal
moduli fields only; (2) Spontaneous breakdown of the permutational
symmetry. Nevertheless, according to ref.[14], it is not easy to find generic
moduli configurations
which realize the latter possibility.

Let us briefly summarize the main  conclusions of \IL [8] concerning
the range of values that can be assigned to modular weights.
Given an orbifold, with point group  in the allowed list of
abelian $Z_N $ or $Z_N \times Z_M$ groups [38], this is associated discrete
sets of twists $\t_i^{(p)},
[p=1,\cdots , T]$ oscillators modings,  \km levels,
and  fixed points
$f_i^{(q)}, [q=1, \cdots , F]$. While Yukawa couplings depend on all these
items, modular weights are essentially fixed by the first three only.
The larger are the values assigned to $k_a$ and to oscillator excitations,
the wider is the  range  of modular weights.
The main issue for \mssm compactification is to
assign modular weights to the
\loe fields consistently  with cancellation  of
duality anomalies and with  gauge \ccs unification.
An important characteristic  of orbifolds here  is the number
of overall unrotated  planes (i.e., planes  left fixed by sweeping through
the entire set of  twists). This
number can take the values: $N_{unr} =3,2,1,0$. The orbifolds
$Z_N\times Z_M$ have $N_{unr}=
3$;  $Z'_6$ has $ N_{unr}=2$; $Z_4, Z_6, Z_8, Z'_8, Z_{12},
Z'_{12}$  all have
$N_{unr}=1$; prime  number orbifolds, $Z_3, Z_7$,  have $N_{unr}=0$.
 The duality  anomalies  cancellation
 places  the constraint,  $ \tilde b_a^{'i}=0$,
for overall  rotated planes, $ i=1,2,3$, where the index
 $a$  labels the \sm group factors, $a=3,2,1$,
corresponding to $SU(3)\times SU(2)\times U(1)$.
For the \mssm the conditions $ \tilde b^{'i}_a =0$
for an overall rotated plane, take the form:  $b_a^{'i} \equiv -c_{a}-
\sum_g c_a^{\a } n_{\a_g}^i = k_a \d_{GS}^i$,
 where $g$ is a  generation index,   $c_{3}=3, c_{2}=5 , c_{1}= 11 $
 and $c^\a _a $  are  rational
numbers,   which  for  \sm fields labeled  in the following order,
$\a =[Q,U^c,D^c,L,E^c, H_1,H_2]$, are given
by [8]: $ c^\a _{3}
=[2,1,1,0,0,0,0],\quad c^\a _{2}= [3,0,0,1,0,1,1],
\quad c^\a _{1}=[ {1\over 3}, {8\over 3},{2\over 3}, 1,2,1,1]$.
These   modular anomalies  cancellation  conditions are absent for the
$Z_N\times Z_M$ orbifolds, because of  $ N_{unr}=3$,
 and involve one equation of the above type (for each $a=3,2,1$)
for the $Z'_6$ orbifold, two for
all other  non-prime $Z_N$ orbifolds and three for prime orbifolds.

The  $\d_{GS}^i$ are extra parameters
which are calculable on a model-by-model basis, for example, by identifying
$\tilde b^{'i}_a=b^{'i}_a -k_a \d_{GS}^i$ to the slope parameter of the
$N=2$ suborbifold  generated by subgroups of the point symmetry
group leaving the $i$-plane fixed.
Since threshold corrections are non-vanishing for
overall unrotated planes only, it follows that the unification constraints
act in complementary manner to the anomalies
cancellation constraints, yielding stronger conditions for
orbifolds with  larger $N_{unr}$.
Of course, it is always possible to partially relax
these constraints by considering
anisotropic orbifolds, since a condition such as
$ T_1\gg T_2, T_3$, which freezes $T_2, T_3$,  renders ineffective
any threshold corrections involving these moduli.
The analysis of \IL [8]  indicates  that  most $Z_N$ orbifolds  fail to satisfy
the
combined anomalies cancellation and gauge unification
constraints, unless one considers higher \km levels
for the non-abelian groups, which is not a
favored option in orbifolds constructions.
 The acceptable canditates, within the minimal framework,  are:
 $Z'_8$,  $Z_2\times Z_3$ and $
Z_2\times Z'_6$ orbifolds. Generalization to
squeezed orbifolds  extends the  acceptable orbifolds to
the whole   $Z_N\times Z_M$ set.
Recall here that anisotropic compactification
could  also represent  a viable mechanism to generate
large hierarchies in the Yukawa couplings matrices [39].
Unlike the superpotential and gauge functions, the
K\"ahler potential is negligibly affected by anisotropies.
Indeed, the constraint $\sum_i \log r_i =0$ removes all dependence here
on $r_i$ at tree level. At one-loop level, a residual contribution may arise
from the term, $\sum_i\d_{GS}^i \log (T_i+\bar T_i)=-
\sum_i\d_{GS}^i \log r_i + \d_{GS} \log (T+\bar T ) $,  which can be
absorbed inside the dilaton field. However,  whereas threshold
effects involve components $\d_{GS}^i$ for overall unrotated
planes, the one loop K\"ahler potential involves all planes, whether
rotated or not.

{\bf 2.2 Soft \susy breaking terms }

The  scalar potential of the N=1 \sugra \sig  is composed of
F- and D-terms,
 $V= V_F+V_D$ [40], defined as:
$$V_F=e^K \big (D_iW(K^{-1})^{i\bar j} D_{\bar j}
W -3 \vert W \vert ^2\big )=(F_i  G^{i\bar j}F_{\bar j}-3e^G), \quad V_D=
\ud D_a (f^{-1})_{aa'} D_{a'} , \eq 7)$$
where $  G=K+\log \vert W\vert^2 $, $D_iW=W_i+K_i W \equiv \dh_i W +W\dh_i
K , \quad D_a=G_i(T^a)_{ij}A_j $,  the moduli and matter fields
are labeled by the index  $i $ and
$ F_i=e^{G/2}(G^{-1})_{i\bar j} G_{\bar j}$
identify with   their  auxiliary  components.
Following the hidden sector approach, we assume that the superpotential
includes a
non-perturbative part which lifts
flat directions in  $S$ and $T_i$ by inducing
a non-trivial   minimum of $V$ characterized by  non-vanishing VEVs
for  dilaton and moduli  scalar $S, T_i$   as well as
auxiliary  field  components $ F_S, F_{T_i}$.
We shall follow the approach of  \BIM  [19] in which
local \susy breaking with $<V>=0$  is assumed to be  saturated
by the  dilaton and moduli
auxiliary fields. Generalizing the parametrization  of
ref.[19],  we represent these quantities
as follows:
$$(G^\ud)_{S\bar S}F_{\bar S}=  \sqrt{3}  m_g e^{i\a_S}\sin \t ,\quad
(G^\ud)_{T_i\bar T_j}F_{\bar T_j}= \sqrt 3  m_g
e^{i\a_T}e_i \cos \t , \eq 8)$$
under the restriction $\sum_ie_i^2=1$,   necessary to ensure $<V>=0$,
and the simplifying assumption that  the anisotropy parameters
$e_i  [i=1,2,3]$ are  real.
The gravitino mass is $m_g=e^{G/2}$ and $\t$ is interpreted
as a dilaton-moduli mixing angle.
The complex phases $\a_S, \a_T$, along with possible  non-vanishing
imaginary parts of VEVs, $Im S, Im T_i$,
represent CP violation parameters.
The isotropic case, which is  considered in ref.[19], can be  recovered by
identifying the three moduli  components, hence
setting $ F_{T_i}= F_T, \quad e_i= 1/\sqrt 3$.

The  case involving  the one-loop  corrected K\"ahler
potential, eq.(5),  can be treated in close analogy  with the above
tree level case by modifying the parametrization as follows [19]:
$${1\over Y} \bigg ( F_S +{2\over (4\pi )^2} \sum_i {\d_{GS}^i F_{T_i}
\over T_i+\bar T_i } \bigg ) = \sqrt 3 m_g e^{i\a_S} \sin \t ,
\quad \bigg (1+{2 \d_{GS}^i\over (4\pi )^2 Y }
\bigg )^\ud {F_{T_i} \over T_i+\bar T_i }=
\sqrt 3 m_g e^{i\a_T} e_i \cos \t . \eq 9)$$
The isotropic case results by setting, $e_i={1\over \sqrt 3}, \d_{GS}^i=
{\d_{GS}\over 3}, \d_{GS}^i {F_{T_i}\over T_i+\bar T_i} =
\d_{GS}{F_T\over T+\bar T}$.
The  parametrization (8)   can also  be generalized to allow
for a matter component of the Goldstino by
introducing an additional mixing angle
$\t_A $ and  writing [19]:
$$\eqalign { (G^\ud F)_{S}=& \sqrt 3 m_g e^{i\a_S} \sin \t \cos \t_A, \quad
 (G^\ud F)_{T_i}= \sqrt 3 e^{i\a_T}m_g e_i \cos \t \cos \t_A,
\cr  (G^\ud F)_{A}=&  \sqrt 3m_g e^{i\a_A} \sin \t_A , \cr }\eq 10)$$
the  one loop  case involving again  a direct generalization of eq.(9).
 Auxiliary and scalar
components  are not independent, of course.
They might be related  by  specifying  the
non-perturbative superpotential. Consider,  for instance,  the gaugino
condensation approach, in which  approximate elimination of the
composite gauge fields associated to a subset $\prod_{A'} G_A'$
of  hidden
gauge group factors, leads to the following toy model
superpotential [13-15],
$$W_{np}(S,T_i)=\sum_{A} h_Ae^{-{3k_AS\over 2b_A}} \prod_i
\big [\eta(T_i)\big]^{\nu_i}, \eq 11)$$
where $ h_A$ are constant parameters and  $\nu_i=-2$ as required
to obtain a modular invariant
function, $G_{np}=
K+\log \vert W_{np}\vert^2$, free of singularities in the $T_i$-planes. By
virtue of
modular invariance, the field space of  the variables $T$
can be restricted to the fundamental region,
$F=[\vert T \vert >1, Im(T)<\ud ] $.
The relation between anisotropy parameters $e_i$
and  the  $r_i$ parameters, defined by $r_iT_i(x)=T(x)$,
can  then be obtained by calculating the auxiliary fields associated with
eq.(11):
$$
(G^\ud F)_{T_i}=m_g\big [ -1+\nu_i  (T_i+\bar T_i) {d\log \eta(T_i)
\over dT_i}) \big ]=
m_g \bigg [-1-{\pi \over 12} \nu_i( T_i+\bar T_i)
\bigg (1-24\sum_{n=1}^{\infty } {n q_i^n\over 1-q_i^n} \bigg ) \bigg ].
 \eq 12)$$
Since for $T_i\in F, \quad  Re(T_i)> \sqrt 3/2,$ and
the moduli dependence involves the variable
$ q_i=e^{-2\pi T_i}$, we see from the size of the exponential
suppression, $q_i< e^{-\pi \sqrt 3} \approx 4.33 10^{-3} $,
that   the relevant
regime   where the  infinite sum  in  eq.(12) needs  to be evaluated is
that of large $T_i$.
Assuming that all  $T_i$  lie in the large compactification radius  regime, one
finds: $
{e_i \over e_j}
\approx {r_j \over r_i}$.
For an isotropic orbifold, the restriction of
equal $r_i=r=1$, likewise implies equal $e_i
=1/\sqrt 3$. For an anisotropic  orbifold, say, with $T_1\gg T_2, T_3$, or
$r_1\ll r_2, r_3$, assuming the large-$T$ limit, leads to:
$e_1\gg e_2,e_3$.

Let us now turn to the implications concerning the soft \susy breaking
interactions. We identify the  string \loe observable sector with
that of the \mssm and assume the presence of a
perturbative superpotential
describing the  renormalizable
trilinear and bilinear Yukawa couplings of
quarks, leptons and Higgs bosons  left chiral superfields,
$$ W=\l^u_{ij}Q_i\e H_2U^c_j+ \l^d_{ij}Q_i(-\e )H_1D^c_j+\l^e_{ij}L_i
(-\e )H_1E^c_j+ \mu H_1\e H_2,  \eq 13)$$
such that $Q_i={u_i\choose d_i} , L_i={\nu_i \choose e_i}
, H_1={H_1^0\choose H_1^-}  , H_2={H_2^+\choose H_2^0} $  denote   quarks,
leptons and
Higgs weak doublets   left-chirality fields, with  the invariant
combination, $ Q\e H=Q_\a\e_{\a \b } H_\b  ,
\quad \e_{\a \b  } \quad   [\e_{12}=1] $  being the antisymmetric symbol;
 $U^c, D^c, E^c $ denote
 antiquarks and antilepton  weak singlets left-chirality   fields  and
the indices  $i, j =[1,2,3] $ label  the squarks or sleptons  generations.
We assume  a one-to-one correspondence with massless string modes,
so that  modular symmetry implies that
Yukawa \ccs $\l_{ij}^x(T_k), [x=u,d,e]  $ transform under $SL(2,Z)_m$
as modular forms of weight,
$(-1-n_{Qi}^m-n_{Uj}^m-n_{H_2}^m) $ for $x=u$ and similar formulas
for other cases. For a given orbifold, these are
calculable  functions of the moduli fields depending on
characteristics of states, such as
oscillator excitations, reference plane
and,   for twisted sector states,
coordinates  of fixeds points [39,41].
Adding the perturbative superpotential  (13) to the non-perturbative one
 and substituting
in the scalar potential, eq.(7), yields  a tree level
soft \susy breaking potential of the familiar form,
$$V_{soft}= \tilde m_\a^2 A_\a A_\a ^{\dagger} +\ud M_a \bar \l_a \l_a +
m_g \bigg [\bigg (A_{ij}^u\l_u^{ij}Q_i\e H_2U^c_j+ A_{ij}^d \l_d^{ij}Q_i(-\e
)H_1D^c_j$$$$
+A_{ij}^e \l_e^{ij}L_i (-\e )
H_1E^c_j+ B_\mu  \mu H_1\e H_2 \bigg )+c.c. \bigg ], \eq 14)$$
with  the following formulas for the soft breaking coupling contants  in
the tree level case[19],

$$\eqalign{ M_a=&  {m_g\over Re(f_a) } \bigg [ \sqrt 3 e^{-i\a_S} \sin \t
({\dh f_a \over \dh S})Re(S) -{1 \over 16\pi^3} e^{-i\a_T}
\cos \t \sum_i \tilde b_a^{'i} Re (T_i) \hat G_2(T_i)\bigg ],\cr
\tilde m_\a^2=&  m_g^2 (1+n_\a{''} \cos^2\t ), \cr
A_{\a \b \g }=& -\sqrt 3 e^{-i\a_S} \sin \t -e^{-i\a_T}\cos \t
\bigg (e'+n'_\a
+n'_\b +n'_\g  -\sqrt 3\sum_i (T_i+\bar T_i) e_i {w^{T_i}_{\a \b \g }\over
w_{\a \b \g }} \bigg  ), \cr
B_\mu =& -1-\sqrt 3 e^{-i\a_S}\sin \t \bigg (1-{\mu_S\over
\mu } (S+\bar S) \bigg ) \cr
&-e^{-i\a_T}\cos \t \bigg (e'+n'_{H_1}+n'_{H_2}-\sqrt 3 \sum_i e_i (T_i+\bar
T_i)
{ \mu_{T_i}\over \mu } \bigg ).\cr } \eq 15)$$
 The index  $\a $ attached to
 \loe  matter fields, denoted generically as
 $A_\a (x) $, subsumes  gauge and generation
quantum numbers  and the
index $a=3,2,1$, attached to gaugino fields, $\l_a (x) $,
labels  the \sm  gauge group factors
$SU(3)\times SU(2)\times U(1)$, in the indicated order.
We have introduced the following notations:
$n^{''}_\a = 3\sum_i e_i^2n_\a^i, \quad
n^{'}_\a = \sqrt 3 \sum_i e_in_\a^i,\quad  e'=\sqrt 3\sum_i e'_i ,
\hat G_2(T)\equiv
G_2(T)- {2\pi \over T+\bar T}, \quad  G_2(T)= -4\pi {\eta'(T)\over \eta(T)}
\approx \pi^2 \big (1/3 - 8 \sum_n nq^n/(1-q^n) \big );$ $w_S =\dh w/\dh S ,
$ with  $w=w_{\a \b \g } A_\a A_\b A_\g$ a generic notation for the
superpotential terms in eq.(13).
The isotropic orbifold case, which is treated in
ref.[19], is recovered by setting,
$ e_i =1/\sqrt 3$, which gives: $e'= 3, n^{''}_\a
=n'_\a = n_\a =\sum_i n_\a ^i, \quad \sqrt 3
\sum_i e_i (T_i +\bar T_i) f_{T_i}
=(T+\bar T )f_T. $
The one loop  improved case  is described by identical
formulas to eqs.(15) with
the substitution:
$$ e_i\to {e_i\over \bigg (
1+{2\d_{GS}^i\over (4\pi )^2 Y }\bigg )^\ud }.$$
The isotropic one loop improved  case is recovered by performing  the
substitutions
described above for the tree level case,
together with  the formal replacement[19]:
$$\cos \t \to \cos \t (1+{\d_{GS} \over 24\pi^2 Y})^{-\ud }, \quad
\sin \t \to \sin \t .$$

{\bf 2.3 Renormalization group analysis}

Distinct  physical theories are parametrized by \ren group trajectories
which trace the scale dependence of the
gauge,  Yukawa and soft
\susy breaking \ccs .
These are described  by first-order
differential equations in the scale variable, $t =\log {M_X^2\over Q^2} $,
with $M_X $ the string unification mass scale and $ Q$ a floating \ren
 scale, so that a solution to these
equations is  fixed uniquely once
boundary  conditions are specified  at $ M_X$, as for  the quantities
quoted in eqs.(15),
or at some other  scale.
While results for the \ren group equations are available in the literature up
to two-loop order, in the present work,
 we shall restrict ourselves to  the
 one-loop equations,  and rely on  refs.[21,23,42] as our main sources
 regarding notational conventions.
The one loop  gauge \ccs   and gaugino masses are then
decoupled from the other parameters. Their   scale dependence
can be expressed explicitly,
 $$ \tilde \a_a(t)\equiv {g_a^2(t)\over (4\pi )^2} =
{\tilde \a_a (0) \over 1- \b_a t +\tilde \a_a (0) \D_a}, \quad
 M_a(t)={M_a(0)\over 1-\b_a t}, \eq 16)$$
 where:
$$ \b_a  \equiv
b_a \tilde \a_a (0) = {b_a g_a^2(0)\over (4\pi )^2 },
 \quad \D_a= \d_a+\sum_i \tilde b_a^{'i} \log \big [(T_i+\bar T_i)
\vert \eta (T_i)\vert^4 \big ]. \eq 17)$$
 The  string threshold
corrections $\D_a $  include a moduli-independent part $\d_a$, to be
neglected in the following, along with the familiar
moduli-dependent part [9]. The
$\b $-function slope parameters $b_a$
are  defined as: $\dh g_a/\dh log Q= -b_a g_a^2/(4\pi )^2$, the values
for the \sm group factors being: $ b_a= (3,-1,-11)$, subject to  the
tree level normalization convention, $g_a^2(0) k_a = g_X^2$.

The \ren  group equations for the Yukawa \ccs  involve  the gauge \ccs and a
dependence on generations through the Yukawa couplings. Motivated by the
large hierarchy  between  third  and first  or second
generations, we shall adopt  here the approximation  where
all entries in $\l^x_{ij}, [x=u,d,e]$  are neglected relative  to the
$(3,3)$ entry. Denoting the corresponding $(33)$  Yukawa matrix elements
as $\l_t, \l_b, \l_\tau $, the index  standing
for top- and  bottom-quark  and tau-lepton, respectively,
one obtains the simplified third generation  evolution equations [23]:
$$\eqalign {{d\tilde Y_t(t)\over dt }=&\tilde Y_t(t)\bigg (  y_{ua}\tilde
\a_a(t)
-6\tilde Y_t(t) -\tilde Y_b(t) \bigg ) ,\cr
{d\tilde Y_b(t)\over dt }=&\tilde Y_b(t) \bigg (y_{da}\tilde \a_a (t)
-6\tilde Y_b(t) -\tilde Y_t(t) -\tilde Y_\tau (t)\bigg ),\cr
{d\tilde Y_\tau (t)\over dt }=&\tilde Y_\tau (t) \bigg
(y_{e a}\tilde \a_a(t)
-4\tilde Y_\tau  (t) -3 \tilde Y_b  (t) \bigg ), \cr } \eq 18)$$
where $ \tilde Y_x(t)={\l_x^2 (t)\over (4\pi )^2}, [x=t,b,\tau ]$ and
$  y_{ua}=({16\over 3}, 3, {13 \over 9}),
 y_{da}=({16\over 3}, 3, {7 \over 9}),
 y_{e a}=(0, 3,  3)$. An analytic solution to
 these equations  exists in the approximation $\l_t\gg \l_b ,
 \l_\tau $ [23],
 $$ \tilde Y_t(t)= { \tilde Y_t(0) E_1(t) \over 1+6\tilde Y_t(0)F_1(t)},
 \quad \tilde Y_b(t)={\tilde Y_b(0) E_2(t) \over
 (1+6\tilde Y_t(0)F_1(t))^{1\over 6}}, \quad
 \tilde Y_\tau (t)=\tilde Y_\tau (0) E_3(t), $$
where:
 $$E_1(t)=\prod_a \bigg ({\a_a(t)\over \a_a(0)}\bigg )^{y_{ta}\over b_a}
 , \quad F_1(t)=\int_0^t dt' E_1(t'), $$$$
 \quad E_2(t)=E_1(t)(1-\b_1 t)^{2\over
 3b_1}, \quad E_3(t)=(1-\b_2 t)^{-{3\over b_2}}
 (1-\b_1 t)^{-{3\over b_1}}. \eq 19)$$

The \susy breaking parameters mix together as well as
with the gauge and Yukawa parameters. For the $A^x_{ij}, [x=u,d,e]$,
we shall adopt   an analogous approximation to the above, namely,
dropping all but the $ (i,j)=(3,3)$ entries. This approximation is
justified by the fact that these quantities  always enter through products,
$A^x_{ij}\l^x_{ij}$. The simplified evolution equations for the $(3,3)$
parameters, designated as $A_t, A_b, A_\tau $, read [23]:
$$\eqalign { {d A_t(t)\over dt }=& y_{ua}\tilde \a_a(t) {M_a(t)\over m_g}
-6\tilde Y_t(t)A_t(t) -\tilde Y_b(t)  A_b(t), \cr
{d A_b(t)\over dt }=& y_{da}\tilde \a_a (t) {M_a(t)\over m_g}
-6\tilde Y_b(t)A_b(t) -\tilde Y_t  (t)A_t(t) -\tilde Y_\tau (t) A_\tau (t),\cr
{d A_\tau (t) \over dt }=& y_{e a}\tilde \a_a (t){M_a(t)\over m_g}
-3\tilde Y_b(t)A_b(t) -4\tilde Y_\tau (t) A_\tau (t). \cr}\eq 20)$$
None of the other matrix elements, $A^x_{ij},
[x=u,d,e]$ with $(i,j)\ne (3,3)$,   are  coupled to the
 $\l^x_{33}(t)$, so that these parameters have  no
other \ren scale dependence in the present
approximation except for that arising from gaugino loops.  This
affects the diagonal elements by a generation independent contribution
analogous to that arising in eqs.(20).
The approximate  solution can be obtained by a simple quadrature,
$$A_{ij}^x (t) = A_{ij}^x (0)+y_{xa}\tilde \a_a(0) {M_a(0)\over m_g}
{t\over 1-\b_a t } \d_{ij}  \quad [x=u,d,e; \quad (i,j)\ne (3,3)]  .$$

The \ren group equations for the scalars masses include contributions
from gauginos loops as well those from scalars and Higgs bosons loops. In the
above approximation of dominant $\l_{33}^x(t)$ Yukawa couplings, the scalars
mass matrix  elements, $(i,j)=(3,3), $ and $(i,3), (3,i), [i=1,2]$
decouple from other  Yukawa matrix elements and have a scale evolution
governed for the $(3,3)$ entries by the equations [21]:
 $$\eqalign { {d(\tilde m^2_{Q})_{33}\over dt}=&z_{Qa}\tilde\a_a M_a^2 -
 \tilde Y_t(\tilde m^2_Q+\tilde m^2_{Q^c}+\bar \mu^2_2+m_g^2A_t^2)
 -\tilde Y_b(\tilde m^2_Q+\tilde m^2_{b^c}+\bar \mu_1^2+m_g^2A_b^2), \cr
 {d\tilde m^2_{Q^c}\over dt}=&z_{Q^ca}\tilde\a_a M_a^2 -
 2\tilde Y_Q(\tilde m^2_Q+\tilde m^2_{Q^c}+{\bar \mu^2_2\choose
 \bar \mu_1^2}+m_g^2A_Q^2),\cr
 {d\tilde m^2_{L}\over dt}=&z_{La}\tilde\a_a M_a^2 -
 \tilde Y_\tau (\tilde m^2_L+\tilde m^2_{\tau^c}+\bar
\mu_1^2+m_g^2A_\tau^2),\cr
 {d\tilde m^2_{L^c}\over dt}=&z_{L^ca}\tilde\a_a M_a^2 -
 2\tilde Y_\tau (\tilde m^2_\tau +\tilde m^2_{\tau^c}+\bar \mu_1^2
 +m_g^2A_\tau^2), \cr }\eq 21)$$
 and for the entries $(i,j) $ with $[i=1,2,\quad  j=3] $ or $[i=3,
 \quad j=1,2]$ by the equations,
$$ {d \over dt } (\tilde m_{Q,Q^c,L,L^c}^2)_{ij}= -\bigg [
\ud(\tilde Y_t+\tilde
Y_b), \tilde Y_{Q}, \ud \tilde Y_{\tau }, \tilde Y_{\tau } \bigg ]
(\tilde m_{Q,Q^c,L,L^c}^2)_{ij}. \eq 22) $$
 We have omitted  writing the argument $t$ for the various
 coupling constants and have denoted $ Q=[t, b],
 Q^c=[t^c , b^c],  L=[\tau , \nu_\tau ] , L^c=\tau^c $
with $ z_{Qa} =4 C_2(R_\a) $, corresponding to the
quadratic Casimir operator  in representation $R_\a $
for the \sm gauge group factor
$a=3,2,1$, such that
$z_{Qa} =({16\over 3}, 3,{1\over 9}),
z_{u^ca} =({16\over 3}, 0,{16\over 9}),
z_{d^ca} =({16\over 3}, 0,{4\over 9}),
z_{L^ca} =(0, 3,1), z_{e^ca}=(0,0,4)$.

For the  scalars  masses matrix elements,
$(i,j)$ with $[i=1,2, j=1,2]$,
we   shall include the contributions of the  Yukawa interactions
perturbatively by
retaining the terms  with leading  power of $t=\log { M_X^2 \over Q^2}$.
The gauge interactions contributions, being generation independent,  have the
same form
as those appearing in eq.(21), and are then included exactly.
The  relevant formulas read [21]:
$$(\tilde m^2_{Q}(t))_{ij}\approx (\tilde m^2_Q(0))_{ij}+\d_{ij}
G_{Qa}(t)
+ t \bigg [\big [-\ud \big (\{\l_u\l_u^\dagger ,\tilde m^2_Q\}\big )_{ij}
-(\l_u\tilde m^2_Q\l_u^\dagger )_{ij}
$$$$-\bar \mu_2^2(\l_u\l_u^\dagger  )_{ij}-m_g^2\bigg ((\l_uA_u)
(\l_uA_u)^\dagger\bigg )_{ij} \big ]
+ [ \l_u \to \l_d , \bar \mu_2\to \bar \mu_1 \big ] \bigg ],$$
$$(\tilde m^2_{Q^c} (t))_{ij} \approx (\tilde m^2_Q(0))_{ij}+\d_{ij}
G_{Q^ca}(t)
+ t \bigg [- \big (\{\l_Q^\dagger \l_Q ,\tilde m^2_{Q^c}\}\big )_{ij}
$$$$-2(\l_Q^\dagger \tilde m^2_{Q^c}\l_Q)_{ij}
-2{\bar \mu_2 \choose \bar \mu_1^2}(\l_Q^\dagger \l_Q)_{ij} -2
m_g^2\bigg ((\l_QA_Q)^\dagger ((\l_QA_Q)\bigg )_{ij}  \bigg ], \eq 23) $$
where we understand that terms on the right hand sides are evaluated
at the \ew scale and where
 $$ G_{Qa}(t)=\ud   z_{Qa}\tilde \a_a(0)
M_a^2(0)f_a(t), \quad [f_a(t)= {(2-\b_a t)t \over (1-\b_at)^2} ] $$
stands for gaugino loop contributions.
The slepton mass matrices $\tilde m^2_L , \tilde m^2_{L^c}$ are
obtained by analogous formulas to those for
$\tilde m^2_Q , \tilde m^2_{Q^c}$ above
by substituting  slepton mass matrix for squark mass matrices
and letting  $ \l_d\to \l_e, \l_u\to 0 $.

 The mass parameters  $\bar \mu_i^2=\mu_i^2-\mu^2  $, appearing in formulas
 above for scalar masses,
 refer to the coupling constants
 in the Higgs bosons  scalar potential,
$$V(H_i)=\mu_1^2 \vert H_1\vert^2
+\mu_2^2 \vert H_2\vert^2 -\mu_3^2(H_1H_2+c.c.)+{g_1^2+g_2^2\over 8}
(\vert H_1\vert^2 -\vert H_2 \vert ^2)^2, \eq 24) $$
where  $ \mu_i^2=\tilde m_{H_i}^2+\mu^2 \quad  [i=1,2] ,
\quad \mu_3^2=-B_\mu m_g \mu $.
For completeness, we quote the scale evolution equations for these
Higgs sector parameters, again in the simplified version in which
our calculations are  ferformed, namely, predominant
$(3,3)$  trilinear couplings [23,42]:
$$\eqalign {  {d\mu^2(t) \over dt}=& \mu^2 \bigg [ 3\tilde \a_2+\tilde
\a_1-3(\tilde Y_t
+\tilde Y_b+{1 \over 3} \tilde Y_\tau )\bigg ],  \cr
 {d\mu_1^2(t) \over dt}=&  3\tilde \a_2M_2^2+\tilde
\a_1  M_1^2 + \bigg (3\tilde \a_2 +\tilde \a_1-3 (\tilde Y_t +
\tilde Y_b +{1\over 3} \tilde Y_\tau )\bigg ) \mu^2 -3\tilde Y_b
(\tilde m_t^2+m_{b^c}^2+\bar \mu_1^2 +m_g^2A_b^2) ,\cr
 {d\mu_2^2(t) \over dt}=&  3\tilde \a_2M_2^2+\tilde
\a_1  M_1^2 + \bigg (3\tilde \a_2 +\tilde \a_1-3(\tilde Y_t +
\tilde Y_b+{1\over 3} \tilde Y_\tau )\bigg ) \mu^2 -3\tilde Y_t
(\tilde m_t^2+m_{t^c}^2+\bar \mu_2^2 +m_g^2A_t^2), \cr
 {d \mu_3^2(t) \over dt}=& -( 3\tilde \a_2M_2+\tilde
\a_1  M_1 -3m_gA_t\tilde Y_t)\mu
+\ud (3\tilde \a_2+\tilde \a_1 -3\tilde Y_t
-3 \tilde Y_b -\tilde Y_\tau )\mu_3^2 . \cr } \eq 25) $$
Let us recall here that the weak hypercharge D-terms add
an extra contribution in the scale
evolution of scalars masses [22],
$${d\tilde m_\a^2\over dt}= -Y_\a \tilde \a_1(t) S(t)+\cdots ,
\quad \bigg  [S(t)={S(0)\over 1-\b_1 t}\bigg ], $$
$$S(t)=\sum_{gen} (\tilde m^2_Q-2\tilde m^2_{U^c}+
\tilde m^2_{D^c}-\tilde m^2_L+\tilde m^2_{E^c} ) -m^2_{H_1} +m^2_{H_2} =
Trace (\tilde m^2_\a
Y_\a ),$$
where $Y_\a =({1\over 6},- {2\over 3}, {1\over 3}, -\ud , 1, -\ud , \ud )$
is the hypercharge of  the \loe  fields  in the order \hfill\break
$(Q, U^c,D^c, L, E^c, H_1,H_2)$,
and dots refer to gauge and Yukawa terms quoted above.
This term  contributes only in cases with non-universal boundary
conditions [22]. Although it  may significantly influence
the spectra,  it was found in ref.[33] to affect
negligibly flavor changing observables and so will be neglected in the sequel.

At the \ew breaking scale, which we shall identify with
the Z-boson mass,  $Q\approx m_Z, \quad t_Z=2 \log{M_X\over m_Z} $,
scalars masses obtain
additional contributions from Yukawa couplings F-terms and
gauge couplings D-terms.
Let us adopt the familiar notation  in which
the three  generations of  left chiral
squarks (L)  and anti-squarks (R) are combined together into a single
six-dimensional coloumn vector, so that the scalars mass terms
in the Lagrangian appear in the  $6\times 6$ matrix form,
$$ -L_{mass}= \sum_{Q=U,D}(\tilde Q_L ,  \tilde Q_R^\dagger )
\pmatrix {M^{Q2}_{LL} & M^{Q2}_{LR} \cr
M^{Q2\dagger }_{LR} & M^{Q2}_{RR}\cr }
{\tilde Q_L^\dagger \choose \tilde Q_R}  +[ Q \to L], \eq 26) $$
The full  squarks and sleptons  masses, comprising the above
\susy breaking contributions evolved from the   the string scale
down to the \ew scale, along with those
induced at the \ew scale,  read [42]:
  $$\eqalign { (\tilde M^{Q2}_{LL})_{ij}(t_Z)=& (\tilde
m^2_{Q})_{ij}(t_Z)+(M_QM_Q^\dagger )_{ij}
  \mp m_Z^2\cos 2 \b
 \big  (\ud -{1\over 3} {2\choose 1} \sin^2\t_W\big )\d_{ij} ,\cr
  (\tilde M^{Q2}_{RR})_{ij}(t_Z)=& (\tilde m^2_{Q^c})_{ij}(t_Z)+(M_Q^\dagger
  M_Q)_{ij}
  \mp m_Z^2\cos 2 \b
  {1\over 3} {2\choose  1} \sin^2\t_W\d_{ij};  \cr
  (\tilde M^{L2}_{LL})_{ij}(t_Z)=& (\tilde m^2_{L})_{ij}(t_Z)+(M_LM_L^\dagger
)_{ij}
  \pm \ud m_Z^2\cos 2 \b
  {1- 2\sin^2\t_W \choose 1} \d_{ij} ,\cr
  (\tilde M^{L2}_{RR})_{ij}(t_Z)=& (\tilde m^2_{L^c})_{ij}(t_Z)+(M_L^\dagger
  M_L)_{ij}
 - m_Z^2\cos 2 \b
  \sin^2\t_W\d_{ij}; \cr }\eq 27.a) $$
  $$(\tilde M^{u2}_{LR})_{ij}=\big [m_g A^u_{ij}  +{\mu \over \tan \beta } \big
]
  (M_u)_{ij}, \quad
  (\tilde M^{(d,e)2}_{LR})_{ij}=\big [m_g A^{(d, e)}_{ij}
  +\mu \tan \beta  \big ]
  (M_{(d,e)})_{ij}; \eq 27.b) $$
  the upper and lower positions in eqs.(27.a) referring to the up-squark
  and down-squark cases, respectively, or to
 slepton and   sneutrino   cases, respectively. We denote by $M_x, [x=u,d,e]$
 the quark and lepton mass matrices, $ M_u=\l_u (t_Z) {v_2\over \sqrt 2},
  \quad M_{d,e}=\l_{d,e} (t_Z){v_1\over \sqrt 2},$ where
 $<H^0_i>=v_i/\sqrt 2 $,  such that
$v_1^2+v_2^2=v^2$, $\tan \beta =v_2/v_1$, $m_Z^2
=(g_1^2+g_2^2)v^2/4$.

The condition of radiative \ew symmetry breaking, as derived by minimizing the
Higgs bosons potential, eq.(24), involves  solving  the  two equations [23,42]:
$$\tan^2\b ={\mu_1^2(t_Z) +m_Z^2/2\over
\mu_2^2(t_Z)+m_Z^2/2}, \quad  \quad \sin 2\b ={ 2\mu_3^2(t_Z)\over
\mu_1^2(t_Z)+\mu_2^2(t_Z)}. \eq 28)$$
For completeness, we also quote
the mass  terms of charginos and neutralinos,
$L_{mass}=-[\chi^{-\dagger }M_c  \chi^+ +c.c. ]-\ud \chi^{0T}M_n\chi^0 ,$ with
$$M_c=\pmatrix {M_2& {g_2v_2\over \sqrt 2} \cr
{g_2v_1\over \sqrt 2} & -\mu \cr  } , \quad M_n=\pmatrix {M_1 & 0&
-{g_1v_1\over 2} & {g_1v_2\over 2} \cr & M_2&
{g_2v_1\over 2} &- {g_2v_2\over 2} \cr  & & 0&\mu \cr & & & 0 \cr }. $$
referring to the basis,  $\chi^\pm =(-i\tilde W^\pm ,\tilde  H_{2,1}^\pm ),
\quad \chi^0 =(-i\tilde B^0, -i\tilde W^0, \tilde H_1^0,
\tilde H_2^0)$.
 The contributions of   super-box or
 super-penguin diagrams  to \fcnc processes, with external quarks and leptons,
 are most conveniently described (in the so-called insertion approximation)  in
terms of
 generation mixing terms for scalars mass matrices in  the
 following representation:
 $$\tilde M^{'2}_{MN}=
 V_M\tilde M^2_{MN} V_N^\dagger, \quad [ M,N=L,R]$$
 corresponding to
 the super-CKM mass bases for the quarks and leptons superfields with
 generation diagonal $D $-terms   Yukawa interactions,
 $\tilde Q^\star_iT^a\l_a q_j \d_{ij} +
 \tilde L^\star_iT^a\l_a e_j \d_{ij}+c.c.$.
 The  transformation matrices $V_M^q  \quad [M=L,R; q=u,d,e]$,
 relating  current  and  mass bases for
 quarks and leptons,  are defined in the conventional way
 so as to satisfy, $M_{q}=V^{q\dagger }_L(M_q )_{diag}
 V_R^{q} \quad [q=u,d,e]$.
The  above notation emphasizes the important fact that flavor changing
observables are sensitive to
\susy breaking as well as to Yukawa interactions.
Note that only
the relative product, identified with the  (Cabibbo-Kobayashi-Maskawa) CKM
matrix $ V=V_{CKM}=  V_L^u V_L^{d\dagger }$,  is  physically observable
if one restricts to fermions weak   interactions, while
additional information  on the structure of  the fermions mass matrices
is needed to  describe the physical flavor changing paramaters.

 {\bf 2.4  Yukawa coupling matrices }

To  obtain boundary conditions for the
Yukawa \ccs  at  the string unification scale,
one might use, as in the so-called bottom-up approach,
the available  experimental information for
quarks and leptons masses and weak  mixing angles
to  determine  the Yukawa  coupling matrices at the
\ew scale and  next evolve  these up to the string  unification
scale by integrating the  \ren group equations [21].
Here, we shall follow a simpler approximate procedure: We
assume  some fixed choice for the Yukawa coupling  matrices
at the unification scale,
with a structure consistent with  orbifolds predictions [39,43],
and   deduce from this the implied form of $V_M^q$ at \ew scale.
The analysis of Casas and Mu\~noz [43]   of the
generation dependence of Yukawa matrices for orbifolds
favors  the following structure,
$$\l^q (M_X)=\pmatrix { 0& a& 0 \cr a& A& c_q \cr
0& c_q& B\cr  }, \quad [q=u,d,e] \eq 29) $$
the entries here being real numbers, a choice  which can always be achieved by
performing  suitable chiral transformation field redefintions. The choice of
eq.(29) has
a suppressed value for the first generation, assumed to arise from
non-renormalizable interactions,  and a nearly diagonal structure for
the second and third generations couplings, assumed to
arise from renormalizable couplings,
the suppression of   off-diagonal terms
being   caused by a proper choice of widely
spaced  fixed points.
The assumed  hierarchy of  parameters in eq.(29) is:
$ B\gg A \gg a , c_q$. For top-quark masses, $m_t\le 100$ GeV, the
scale evolution of the Yukawa matrix is dominated by gauge bosons loops
and can be represented  approximately
as an overall scaling factor  determined by the gauge coupling constants,
$$M_{q}(t)= F_q(t)M_q (0) ,\quad
F_q(t)=\prod_a \big ({\a_a (t)\over \a_a(0)}\big  )^{y_a/b_a} . $$
Full formulas for the scale evolution factor are quoted in ref.[43].
For larger top-quark masses, the  top-quark Yukawa couplings
compete with gauge interactions and affect mainly the matrix
elements, $\l^x_{ij}$ with  $i=3$ or $ j=3$, corresponding here to
the parameters $ B(t) $ and
$c_q(t)$.
Considering, for definiteness,  the down-quarks case,
the  mass eigenvalues for a  Yukawa matrix of the  form  of eq.(29) are:
$ m_d\approx a^2/A, m_s\approx A , m_b\approx  B $, so that only $m_b$ is
scale-dependent in the approximation of dropping Yukawa
interactions in the scale evolution, while $c_q$ is scale dependent if
one includes third generation couplings.
For symmetric  mass matrices, as is the case here,
left and right transformation matrices are equal,
$V^d_L=V^d_R$, and given by:
$$V^{d}_L=V_R^{d} \approx \pmatrix { 1& -\sqrt{m_d\over m_s}& {\sqrt {m_d m_s}
c_d
\over m_b m_s} \cr
 \sqrt{m_d\over m_s} & 1 &-{c_d\over m_b} \cr
{\sqrt {m_d m_s} c_d
\over m_b^2  } &
{c_d \over m_b}&  1 \cr }, \eq 30) $$
so that the    transformation matrices are fully specified, up to a single
unknown parameter,
$c_d=c_d(t_Z)$ [43].  Note that $(M_d)_{diag}=(-m_d, m_s, m_b)$,
and that the negative mass eigenvalue can always be absorbed, if so
desired, by a chiral
transformation, say,  $d_L\to -d_L, d_R\to d_R$.
We have only quoted above explicit formulas for the
down-quarks case.
We shall assume that  similar structures hold  for
up-quarks and leptons mass matrices, so  one obtains  for these cases
analogous formulas
by replacing $(d,s,b)\to (u,c,t) $ or $ (e,\mu ,\tau ) $  with
two  additional free parameters, $ c_u(t), c_e (t)$.
The values of these  parameters
at the \ew scale can be obtained  by identifying $ V_L^uV_L^{d\dagger } $
with the CKM-matrix, which supplies four relations  expressing
$ c_u(t), c_d(t) $ as linear combinations
of the CKM matrix elements,
 $ V_{ub}, V_{cb}$ or $ V_{td}, V_{ts}$.
We shall use in our  calculations the above form of Yukawa couplings
for up and down  quarks and for leptons,
and   following ref.[43], tentatively   set the $c_q(t)$ parameters
as follows: $c_u(t_Z)\approx m_s\sqrt 2, c_d(t_Z)\approx m_c,
c_l(t_Z)=m_\mu $.
Of course, other choices  for the Yukawa matrices
will lead to
alternative expressions for the transformation  matrices.
For symmetric Yukawa matrices, under restrictive assumptions on the number of
texture zeros,
the number of  choices is known to be  limited to only  five cases [44].

{\bf 3. Results and discussion}

In this section we shall present numerical results for the
dimensionless parameters defined by:
$$\d_{MN} = {\tilde M^{'2}_{MN}\over \tilde m^2 } =
{V_M^q\tilde M^{q2}_{MN}V_N^{q \dagger } \over
\tilde m^2}, \eq 31) $$ where the mass factor in the
denominator $\tilde m $  represents
some average scalar superpartner mass,  which
we shall set for  the equal chirality cases
$\d_{MM} $ to   the weighted trace of the  mass
submatrix of fixed chirality $ M=L,R$ and for the mixed chirality case
$\d_{LR}$ to the
weighted average of the full mass matrix trace.

{\bf 3.1 Inputs }

The basic parameters entering the  calculations are set as  follows:

{\it (a) Masses:} $m_Z=91.17 GeV, v=246 GeV ; m_t=170
GeV, m_b=5.6 GeV, m_\tau= 1.784 GeV  ,  m_c=1.35 GeV,
m_s=0.175 GeV, m_\mu= 0.1056 GeV ,
m_u= 5.1 MeV, m_d= 8.9 MeV  , m_e=0.51MeV   . $

{\it (b) Gauge coupling constants :} $ g_1^2(m_Z)=0.127, g_2^2(m_Z)=0.425,
g_3^2(m_Z)=1.44 , g_X={1\over \sqrt 2},
M_X = 5.27 g_X 10^{17} GeV =3.73 10^{17} GeV , $ with $ t_Z=2 \log {M_X\over
m_Z} =71.9$.
The boundary conditions  $\a_a (0)$ are evaluated  via eq.(16).

The dilaton VEV is set at  $<S> = {1\over g_X^2} =2$. The numerical
values for the moduli VEVs
will be chosen in the various  cases to be
considered below  on the basis of fits to  gauge \ccs
unification, and specifically to $\a_s (t_Z) $ and $\sin^2 \t_W (t_Z)$ [8,45].
Recall that the radiative corrections to these quantities involve the
linear  combinations, $A^{'i} =
{k_2\over k_1} b^{'i}_1
-b_2^{'i}, \quad B^{'i}= b_1^{'i}+b_2^{'i} -{k_1+k_2\over k_3} b_3^{'i}$,
so defined as to be
free of the $\d_{GS}^i$.
Note that the typical values of
order $ T\approx 10$  that are found in these fits [8] lie
an order of magnitude above those found in mimimizing the
scalar potential
in gaugino condensation models [17]. In the present work
we shall not consider CP violation effects and therefore will set
to zero  the imaginary parts of  dilaton and moduli VEVs and  phases
of auxiliary fields, $\a_S=\a_T =0$.
We also shall  neglect contributions to  \susy breaking  parameters, eq.(15),
involving
derivatives \wrt dilaton and moduli fields of the Yukawa coupling constants.

The Green-Schwarz parameters  $\d_{GS}^i$, representing  gauge group
independent contributions to duality anomalies cancellation, are calculable for
orbifolds on a model-by-model basis via the  $\beta$-function  slopes of
$N=2$ suborbifolds with $i$-plane fixed. The typical values in a minimal \sm
orbifold can be  calculated from formulas given  at the end of Subsection 2.1.
One finds for  the  diagonal sums,
 $ \d_{GS}=\sum_i \d_{GS}^i $, values of positive sign inside the interval,
  $\d_{GS} \approx [5-10] $. (Note that our conventions for the
  parameters $b_a, b^{'i}_a, \d_{GS}^i $ differ by a sign
  from those of ref.[19].) This contrasts with the range  $\d_{GS} \approx
[25-50]$, favored by gaugino condensation models [17].

The parameters $ b^{'i}_a$ also are model dependent. If large  moduli VEVs are
used, then, because of the exponential dependence on $T_i$, this
strongly  constrains the  $\tilde b^{'i}_a$ of  overall
unrotated planes  if one is  to avoid excessive threshold corrections. Indeed,
expressing these  corrections to gauge \ccs in terms of an effective string
unification
scale $M'_X$, one finds:
$${M'_X\over M_X}=\prod_i \bigg [(T_i+\bar T_i) \vert
\eta(T_i)\vert ^4 \bigg ]^{-{\tilde b_a^{'i}\over 2b_a^i}} \approx
\prod_i (T_i+\bar T_i)^{\tilde b_a^{'i}\over 2b_a^i}
e^{\pi T_i\tilde b_a^{'i}\over 6b_a^i} , \eq 32) $$
where the large radius limit was used in the second step.
We see that a reduction of order $10$ in the string   scale  requires
for the exponent  $-{\tilde b^{'i}_a\over 2b_a}$ a positive value
of absolute magnitude close to unity.
An analogous constraint applies for the more relevant exponent,
$-{\tilde b^{'i}_a k_k-\tilde b^{'i}_b k_a \over 2(b_ak_b-b_b k_a)} $
associated to the unification of
two group factors, $G_a  $ and $ G_b$.

The  experimental information
for $\d_{MN}$  can be
obtained  from
\fcnc  observables  involving mass differences in
the neutral mesons systems ($K-\bar K, B-\bar B, D -\bar D$)
and branching ratios of transitions such as
$q_i\to q_j + \g  $ or $ l_i\to l_j + \g , \quad [i\ne j] $. Upper bounds are
derived by assuming that  these observables are solely accounted
in terms of contributions of superbox or superpenguin diagrams
with gluino and squark, or photino and  slepton,  exchanges.
Denoting the average superpartner scalar mass by
$\tilde m$,  the  gluino and  minimal  neutralino masses    by
$M_3$   or $M_{\tilde \g }$, then the  calculated  mass differences
$\d m_K , \d m_B, \d m_D $ and branching ratios $B$  have
a dependence on superpartners masses which can be separated into two variables:
$\tilde m$ and $ x_{3,\g }={M_{3, \g }^2\over \tilde m^2}$. A study of formulas
quoted in refs.[20,34] reveals that
$\d m_{K,B}\propto {\d_{MN}^2 \over \tilde m^2} F_B (x)$, with $F_B(x) $
smooth functions of $x$, so that
the  box diagrams bounds
imply the scaling law,
$ \d_{MN}  \propto  \tilde m, \quad  [M,N = L,R]$.
Using $x=1$ as a reference value, then account of the $x-$dependence  results
in stricter (weaker) bounds depending on $ x<1 (x>1)$.
The penguin diagrams contributions to branching ratios,
neglecting the  contributions of $ A-$ and $\mu -$ coupling terms [20], read:
$B\propto \d_{MM}^2F_{P1}(x)/
\tilde m^4 $ or $ \d_{LR}^2 F_{P2}(x)/\tilde m^2$, which
implies the scaling laws,
 $\d_{MM} \propto  \tilde m^2 $,
 $\d_{LR}\propto  \tilde m$, times smooth functions of $x$.

The numerical  values of  presently known  bounds for
$\d_{MN}$, as obtained from  results in refs.[20,27,34],
are quoted in Table 1.
More stringent bounds are found for  the quantities in Table 1  designated
as, $\d_{ij}=(\d_{LL}\d_{RR})^{\ud }$, which correspond to contributions
implying simultaneously  non-vanishing chiral  left and right mass terms,
$\d_{LL} \d_{RR}\ne 0$. A similar situation holds for
the case of simultaneously  non-vanishing $(\d_{LR})_{ij} $ and  $
(\d_{LR})_{ji}$ matrices.  The bounds in Table 1 for $\d_{LR}$  correspond
precisely to the geometric average of these matrix elements.

{\bf 3.2 General features of model}

It is useful to identify the various sources of off-diagonal contributions
to the matrices $(\d_{MM})_{ij}$. The F- and D-terms (second and third
terms in the formulas of eq.(27.a)) yield purely diagonal contributions.
The solutions of the \ren group equations
(first terms in eq.(27.a)) give contributions the scalars mass matrices with
the approximate structure, $(\tilde M^{Q2}_{LL})_{ij} (t) \approx
(\tilde m^{Q2})_{ij} (0) +G_Q(t_Z) \d_{ij} +\sum_{q'=u,d} y^{(q)}_{q'}
(M_q M^\dagger _{q'})_{ij} $, under the simplifying assumption
that  the contributions  from boundary conditions,
gauginos couplings   and Yukawa couplings
add up linearly. Off-diagonal matrix elements  cannot
arise from the family independent  gauginos  couplings. They also
cannot arise from the boundary conditions  mass matrices if these feature
family universality ($\tilde m^2_{ij}$ is  multiple of identity) or
alignment  with the  fermions  mass matrices
($ V_M \tilde m^2_{MN} (0) V_N^\dagger $ are diagonal matrices).
As is also  well known  [20], the radiatively induced
off-diagonal terms  from Yukawa
couplings are numerically significant only for
the submatrix $ \tilde M^{d2}_{LL} $, yielding for flavor changing parameters
the approximate result:
$$(\d_{LL}^d )_{ij}\approx {y_u^{(d)}\over \tilde m_d^2}\bigg ( V^\dagger
(M_u^2)_{diag }
V \bigg )_{ij}
\approx y_u^{(d)}{m_t^2
\over \tilde  m_d^2} V^\dagger _{i3}V_{3j}, \eq 33)$$
with $y^{(d)}_u$ of order unity and $ V=V_{CKM}$.
(The chirality $L$ is singled out because only
left-chirality scalars of given charge interact with scalars of both
charges and the charge $d$ because of the predominant  top-quark
Yukawa coupling.)
For these flavor changing
terms of \sm origin one expects then  important suppression due to
small CKM mixing angles.
Using numerical estimates quoted in ref.[34] for $y_u^{(d)}$,  gives:
$ \d^d_{12}\approx 7.3 \times 10^{-4}, \d^d_{13}\approx 1.4 \times  10^{-3}$.

For the mixed chirality parameters, the relevant
terms in the interaction basis which  contribute to
off-diagonal elements of $(\d_{LR})_{ij}$  can be written  approximately as:
$$(\tilde M^{2x}_{LR})_{ij} \approx m_g A^x_{ij} (t_Z)M^x_{ij} (t_Z)
\approx M^x_{ij} (t_Z) \bigg [-m_g \cos \t (n_i+n_j) +M_a(0) y_{xa}\tilde \a_a
(0)
\d_{ij} {t\over 1-\b_a t} \bigg ]. \eq 34) $$
This formula  clearly illustrates the fact that non-universal  contributions
(first term) could  be masked
by gaugino contributions (second term),in the  presence of
large gauginos masses.
Because of the dominant $(3,3)$ entries in the fermions
mass matrices, the expected pattern is: $(\d^x_{LR})_{23} >(\d^x_{LR})_{13}
> (\d^x_{LR})_{12} .$

Returning now to the unmixed chirality case,
a simple  estimate   of the generation non-universality of boundary conditions
can be performed  by  including the contributions of gaugino loops in the scale
evolution only, while  dropping entirely those of  Yukawa couplings.  The
physical mass matrices in the weak interaction  basis are then  diagonal
ones and given by:
$$(\tilde M^{x2})_{ij}(t_Z) = m_g^2 \bigg (1 +n_{xi}\cos^2 \t +{3\over 2}
\sin^2 \t \sum_a z_{xa} \tilde \a_a (0) f_a (t_Z) \bigg ) \d_{ij},\quad [x=Q,e]
\eq 35) $$
using self-explanatory  notations, with $f_a(t)$  defined at eq.(23) and
dropping momentarily the suffix $MM=LL, RR$.
Performing the transformation to the super-CKM basis  under the simplifying
assumption of negligible off-diagonal matrix elements
$\tilde M^2_{ij} \approx 0 \quad [i\ne j]$, one obtains
the  approximate formulas:
$$ \tilde M^{'x2}_{ij}=( V^x_M\tilde M^{x2} V^{x\dagger }_M)_{ij} \approx
\tilde M^{x2}_{ii}
V^{x\dagger }_{Mij}+ \tilde M^{x2}_{jj}V^x_{Mij}+
(\tilde M_{33}^{x2}-\tilde M_{11}^{x2} )
V^x_{Mi3}V_{M3j}^{x\dagger }+ \cdots , \quad [x=Q,e]$$
where dots refer to contributions proportional
to $(\tilde M^{x2}_{ii} -\tilde M^{x2}_{22}). $
Given the structure of $V_M$, eq.(30), the expected pattern for the
flavor changing  contributions associated with the boundary conditions is:
$\d_{12}\approx \d_{23}>\d_{13} $.
Neglecting the F- and D-terms,
one obtains for $(i,j)=(1,2) $ or $(1,3)$  the rough  estimate:
$$ \d^x_{ij}\approx 2 (V^{x\dagger }_M)_{ij}
 {\tilde M^{x2}_{ii} -\tilde M^{x2}_{jj} \over
\tilde M^{x2}_{ii} +\tilde M^{x2}_{jj} }  \approx  { \cos^2\t \over 19}
(V_M^{x\dagger })_{ij} {n_{xi}
-n_{xj} \over 1+{1\over 19} \big ({n_{xi}+n_{xj}\over 2}-18\big )
\cos^2 \t },  \eq 36)   $$
which shows that gauginos can  have a strong dilution
effect on non-universality
at the unification scale, to the extent that
the  mixing angles $\t $ lie not too
close to $ 0 $ or $ \pi $.

We shall now focus on the right chirality d-squarks case since this
possesses simplifying features allowing to find upper bounds on the flavour
changing  parameters.
Only the diagonal matrix elements  here get renormalized by gauginos loops.
Since this is proportional to the identity matrix it follows
that the right chirality d-squark mass matrix has non-renormalized
non-diagonal terms.
Let us write down  this
matrix at the compactification scale in the basis where
the d-quark  Yukawa couplings are diagonal:
$$\tilde M'^{Q2}_{RRij}={m_0^2\over 3}\bigg (\delta_{ij}
+\hbox{cotan}^2 \theta \sum_k V^d_{Rik}
(n_k+1)V^{d\dagger}_{Rkj} \bigg ),  \eqno (37)$$
where $m_0=M_a(0)$ is the  common gaugino mass  at unification scale
and  we denote, for notational convenience, $n_i =n_{Di}$.
Defining the renormalized average mass by  means  of  a
numerical estimate for the  gaugino loops contributions,
$$\tilde m^2_{av}={m^2_0\over 3}(18 +{{n_1+n_2+n_3+3}\over
3}\hbox{cotan}^2{\theta}), \eqno(38)$$
we find the flavor changing parameters,
$$(\delta_{RR}^d)_{ij}= {{\sum_k V^d_{ik}n_kV_{kj}^{d\dagger}}\over
{19\tan^2 \theta +{(n_1+n_2+n_3+3)\over 3}}}. \eqno(39)$$
This ratio  explicitly depends on the arrangement of modular weights.
Using  the ansatz for the $V^d$ matrix, eq.(30), the numerator can be
evaluated. As we are looking for an upper bound we choose the arrangement
generating the largest contribution. We also safely neglect the third
family angle compared to the Cabibbo angle. Using unitarity the result
can solely be expressed as a function of modular weight differences.
Moreover the $\theta$ angle can be constrained by requiring that the
scalar masses
are positive at the compactification level. This implies that
$\tan^2\theta>-{1\over (n_m+1)},$
where $n_m$ is the lowest modular weight. This yields the  upper bounds:
$$\eqalign{
(\delta_{RR}^d)_{12}&\le \sqrt{{m_d\over m_s}}{{\Delta n_1}\over
{-19(n_m+1)+{{\Delta n_1+\Delta n_2}\over 3}}},\cr
(\delta_{RR}^d)_{13}&\le{{\sqrt{2m_dm_s}}\over m_b}{{\Delta n_2}\over
{-19(n_m+1)+{{\Delta n_1+\Delta n_2}\over 3}}}\cr
(\delta_{RR}^d)_{23}&\le \sqrt 2 {m_s\over m_b}{{\Delta n_2}\over
{-19(n_m+1)+{{\Delta n_1+\Delta n_2}\over 3}}},\cr}
 \eqno(40)$$
where $\Delta n_1$ is the modular weight difference involving
$n_1$, e.g., $\Delta n_1=n_1-n_3$ if $n_m=n_3$, and similarly for
$\Delta n_2$. Notice that the order of magnitude of these mass
insertions
is given by the ratio of d-quark masses. Effectively these ratios are
not very different from the elements of the CKM matrix. Indeed the
ratios of the u-quark masses composing the u-quark Yukawa coupling
diagonalising matrix are negligible compared with their d-quark
counterparts. Thence $(\delta_{RR}^d)_{12}$ is proportional to a very good
approximation to $\sin \theta_C$. Numerical values can be extracted
from eq.(40). For instance, using  $n_1=-2, \ n_2=-3,\ n_3=-1$,  one obtains
($\delta_{RR}^d)_{12}\le 6.10^{-3},\ (\delta_{RR}^d)_{13}\le 2.5\ 10^{-4},\
(\delta_{RR}^d)_{23} \le 1.1\ 10^{-3}$.
These analytic bounds are modified numerically by the subleading
Yukawa couplings. Nevertheless their orders of magnitude is retrieved.

{\bf 3.3 Non-universal modular  weights assignments }

Most  examples of solutions quoted  by \IL [8]
are generation independent. It is clear, however, that
relaxing the generation
independence constraint could only increase the number
of solutions.
Rather than repeating the type of analysis in ref.[8] to seek for generic
features of solutions  involving a generation dependence,  we choose
here to consider three representative solutions, taken among the
examples quoted in ref.[8],  and consider generation dependent
discrete perturbations of \mow  by a few units.
We do not require  consistency  with  anomalies  cancellation and unification
constraints. In all cases discussed below, we set the \sm \km levels at:
$k_a=(1,1,{5\over 3})$.
The following  three cases will be considered:

(1)  Case I:
(ref.[45] and ref.[8],  eq.(4.6))
This refers to a general, unspecified orbifold
case  with one modulus,  constrained by gauge unification only
and  by a restricted  interval  for \mow :
$-3 \le n_\a \le -1$. It can be realized in terms of an isotropic
orbifold  or  an anisotropic one  with
$ T_1\gg T_2,T_3$, both cases  with all planes  overall unrotated, $N_{unr}=3$,
or with two rotated planes. The generation
independent  isotropic  case, disregarding therefore
the anomalies  cancellation condition, admits an unique solution consistent
with the unification conditions on  the  strong and weak angle  \ccs  [45]:
$n_Q=n_D=-1, n_U=-2, n_L=n_E=-3,
n_{H_1}+n_{H_2}=-5,-4$, with $Re(T)\approx 16 $.
(The suffices $U, D, E$  here obviously  refer to the corresponding
antiparticles.)
We  choose  here, $n_{H_1}= -2,  n_{H_2}=-3$.
Using the above values of  \mow  yields:
$b'_a=(6,8,18)$. The exponential dependence of threshold corrections
and large VEV  of the overall modulus  entail a careful tuning of $\d_{GS}$. We
shall set
$\d_{GS}=7 $, i.e., $ \tilde b^{'i}_a=(-1,1, {19\over 3} $,
which gives ${M'_X\over M_X}\approx (10^{-1},
10^{-3} , 10^{-2}) $ for $[a=3,2,1]$.
We shall implement   generation dependence
in terms of two  subcases:  (i) Case I-a:  $n_{\a i} = n_\a-1 ,
n_{Li}=n_L+1 , n_{Ei}= n_E+1, [i=1,2] $ and
$ n_{\a 3} =n_\a , n_{L3}=n_L, n_{E3}=n_E  $ with $\a = Q, U^c, D^c$ ;
(ii) Case I-b: Same assignments with $Re(T)=1.3$.

(2) Case II: (ref.[8], Section 4)   $Z'_8$ orbifold with two rotated planes,
$i=2,3$, consistent with unification, $Re(T_1)\approx
 24 $ and anomalies cancellation, $\tilde b^{'2,3}_a=0$.
 This admits several generation dependent solution, of which an example is:
 $n^i_{Q123}=(0,-1,0), n^i_{D123}=(-1,0,0),
 n^i_{U1}=-\ud (0,1,1), n^i_{U23}=-{1\over 8} (6,15,3),
 n^i_{L1}=-{1\over 8}(14,3,7), n^i_{L23}=-{1\over 8} (14,7,3),
 n^i_{E1}=-(1,0,0), n^i_{E23}=-{1\over 8} (6,15,3),
 n^i_{H1}=-{1\over 4}(4,3,3), n^i_{H2}=-{1\over 8} (14,3,7)$.
 The $\b $-functions slopes are calculated to be: $b^{'1}_a= (
{3\over 2},{5\over 2},{15\over 2}), \quad  b^{'i}_a= k_a \d_{GS}^i=
{1\over 4}{29 \choose -7}, $ for $i={2\choose 3}$. We shall choose $\d_{GS}^1
={5\over 2}$,  which gives $\d_{GS}= 8$.
The anisotropic torus  with $ T_1\gg T_2, T_3 ,
\quad e_2,e_3 \ll e_1 \approx 1$  has the
modified  weights, $n''_\a  \approx \sqrt 3 n'_\a $, defined at
 eq.(15), given as:
 $n''_{Q123}=0, n''_{D123}=-3, n''_{U123}=(0,-{9\over 4},-{9\over 4} ),
 n''_{L123}=-{21\over 4},  n''_{E123}=(-3,-{9\over 4},-{9\over 4} ),
 n''_{H_1}=-{3\over 2},  n''_{H_2}=-{21\over 4}$. We see that these
effective  weights cover a wider
range than in the isotropic case.
We consider two versions  based on this example:  (i) Case II-a: Isotropic
torus with overall weights:
 $n_{Q123}=n_{D123}=-1, n_{U1}=-1, n_{U2,3}=
-3, n_{L123}=-3, n_{E1}=-1, n_{E2,3}=-3, n_{H_1}=-2,
n_{H_2}=-3$.
(ii)  Case II-b: Anisotropic torus  with $ T_1\gg T_2, T_3 ,
\quad e_2,e_3 \ll e_1 \approx 1$.
To simplify calculations we represent this case by using the usual weights,
but with values covering a wide interval of the type described above. We use:
 $n_{Q123}=n_{D123}= n_{L123}=(0,-2,-3), n_{U123}=n_{E123}=(-1,-3,-3),
n_{H_1}=-2, n_{H_2}=-3$.

(3)  Case III:  (ref.[8], Section 4) $Z'_6$ orbifold with
one rotated plane, $i=3$, consistent with unification, $
Re(T_1)=Re(T_2)\approx 10 $, and anomalies cancellation.
The first condition yields  $ \sum_{i=1,2}
b_a^{'i}=(5,4,-{14\over 3})$  and the second yields $b^{'3}_a =
k_a \d_{GS}^3= -2$. To achieve  satisfactory  gauge \ccs  unification
we shall choose: $\sum_{i=1,2}    \d_{GS}^i =6$.
Starting with the  universal solution of ref.[8],
$n_{Q}=n_{D}= n_{U}=
n_{L}=n_{E}=-1, n_{H_1}=n_{H_2}=-1, $
generation non-universality is implemented by considering two subcases.
(i) Case III-a: $Re (T_1 )=10 $,  $n_{\a1}=n_\a ,
n_{\a 2}=n_\a -1 , n_{\a 3}=n_\a -2 $; (ii) Case III-b:
$n_{\a 12}=-1, n_{\a 3}=-2 , \quad \a =[Q, U^c , D^c, L, E^c]$.

{\bf 3.4 Electroweak  breaking,  soft parameters and \loe spectrum}

We  choose the   dilaton-moduli mixing angle  $\t $  and gravitino mass
$m_g$  as  our sole  free parameters and determine the remaining parameters,
namely, $\tan \b $ and $ \mu $,   by solving eqs.(28) at the \ew scale with the
solutions of the \ren group equations
for $\mu_i^2(t), \quad [i=1,2,3]$ obtained by integrating the
coupled equations (20), (21), (25), using  eq.(19) for Yukawa couplings.
We shall present detailed results
for the following choice of \susy  breaking
free parameters: $m_g=200 $ GeV, $ \t \in [0,\pi  ]$
and for Yukawa \cc : $\l_t(t_Z)=0.85$.  Solving the minimization
equations (28) to find $\tan \b $ and $\mu $, will then  determine all
remaining parameters of the model.
Let us first  try to assess  the accuracy of the
approximation $\l_t\gg \l_b, \l_\tau $ that we shall use. This is expected to
be
a reliable one provided $\tan \b $ is of order unity. Thus,
starting, say,  with $\tan \b =5.5 $  and $\tilde Y_x (t_Z)=[5.8 10^{-3}, 2.0
10^{-4},
6.4 10^{-5}], \quad [x=u,d,e]$ and integrating upwards to large  scales in
the above approximation, yields:
$\tilde Y_x (0)=[3.7 10^{-4}, 1.2 10^{-5},
7.7 10^{-6}]$.  Integrating now downwards  from $M_X$
with the exact evolution equations,
yields:
$\tilde Y_x (t_Z)=[3.1 10^{-3}, 1.6 10^{-4},
2.2 10^{-5}]$, which is in fair agreement with the starting values.

The numerical applications  presented in this work  should usefully
complement those of ref.[19]. The Calabi-Yau and  O-II (orbifolds with small
threshold corrections) scenarios have no impact on the \fcnc issue.
Building up on the  (large threshold correction) O-I scenario  of
ref.[19], we consider here
a wide variety of non-universal modular  weights and also
include the  b-quark and $\tau $-lepton  Yukawa couplings.
The latter item opens up the possibility of independently fixing $\tan \b $
when solving for the radiative \ew breaking condition.
However, the approximation $\l_t \gg \l_b, \l_\tau $, eq.(19),
that we use is a limiting factor here, because this weakens  the
sensitivity  to these \ccs of the
\ew symmetry breaking equations.
We should  mention at this point  that our
method of solving for $\tan \b $ is not very robust.
We proceed numerically  along the following steps: First,  we
select  a trial value
for $\tan \b $,  which fixes $\l_b (t_Z) , \l_\tau (t_Z)$. Second,
we evaluate all other parameters by solving
the \ren group equations for
a discrete set of values for ${\mu (0)\over m_g }$ inside the interval $
[1-4]$.  Finally, we  solve for $\mu  (0)$  the combination of the two
equations (28) obtained by elimination of the explicit dependence on
 $ \tan \b $; we  determine the associated value  of $\tan \b $   and run
down again  the scale evolution equations with these up-dated values
of $\mu (0) $ and $ \tan \b $.
While this procedure leads to accurate solutions
for the second equation (28), thanks to  its strong sensitivity to $\mu (0)$,
it  fails in general   to  accurately solve  the first equation.
The inaccuracies originate from the disparity between the two
scales in the problem,  $m_g $ and $m_Z$, and
worsen with increasing $m_g$. For
$m_g\le 200$ GeV, they  represent from  a few $10 $ \%  in favorable cases to
$50 $ \%. However, the inaccuracies dramatically increase
with larger values of  $m_g$.

Results are shown in figure 1 for case I-a and figure 2 for case III-a. We
see (figures a-b) that solutions with
$\tan \b $ of order unity and concomitantly
$\mu  (0)$  large in comparison to $m_g$,   are generally selected.  The
variation  with $\t $  of these solutions to the \ew breaking equations
depend on the boundary conditions used for Higgs bosons masses. For widely
unequal Higgs bosons  masses at $M_X$  (cases I or II), with increasing $\t $,
$\tan  \b $ decreases
and $\mu $ increases; the opposite takes place for equal masses  at
$M_X$ (case III).
The physical (scale $m_Z$)  masses of  gauginos, scalars  and
fermionic  superpartners  (figures (c), (d), (f))
are all increasing functions of $\vert
\sin \t  \vert $. For gaugino masses, this is due to the fact that
the dilaton term dominates over the moduli term. While the latter
term increases with $<T>$ and with $\d_{GS}$, in all cases that we
deal with here, it represents a small, few \% fraction of the former
and is of negative sign. Note that the moduli
contribution is what prevents mass ratios of the three gaugino masses from
being $\t$-independent.

The scalars mass matrices in the interaction basis
are diagonal ones, due to our restriction to
canonical kinetic terms and the simplified version used for the
scale evolution. The values  of the  averaged traces of the   scalars
mass  matrices are  seen to lie at a factor $2-4$ $ (1)$  times  the basic
energy scale, $m_g$ for squarks (sleptons).  The third
generation masses only are affected by the  Yukawa couplings. These
contributions shift down
the  $(3,3)$  scalars masses in comparison to   the $(1,1), (2,2)$
masses by  a few \%
for squarks and $10-20\% $ for sleptons. (Since we do not exactly
diagonalize the full mass matrices, by taking  into account
left-right chirality  mixing, the foregoing statements are approximate ones.)
The top-quark and  Higgs boson  masses, which have not been plotted
in the figures, are also smooth
increasing functions of $\vert \sin \t \vert $. As
$\t $ sweeps the interval $[0,\pi ]$,
the top-quark   mass in case III, for instance,  increases monotonically in the
interval
$ [111-  150] $ GeV; the corresponding interval for
the tree level  Higgs boson mass is $[12- 90]$ GeV.
For case I, the variation with $\t $ is of opposite sense.
The  third generation trilinear parameters $A^x(t_Z)
$ (figure (e))  are strongly
influenced by radiative corrections.
For case I, as $\t $ sweeps the interval $[0,\pi ], \quad A_t (0)$ regularly
decreases from $1.4 $ to $-3.4$, while $A_t(m_Z)$ (figure  1 (e))
first levels off  then  slowly
decreases from $4$ to $ 0.6$. The parameters $A_{b,\tau }$ are less
affected by radiative corrections.
This is also the case for the $B-$parameter,  for which
$B(0)$  regularly decreases from $-0.4 $ to $-3.6 $ while
$B(t_Z)$ decreases  from $-0.3$ to $-2.5$.

The rapid decrease with decreasing $\t $ of the average scalars masses
near small $\t $ or $  \pi -\t  $ is explained by
the fact that the   masses of scalars    with finite
\mow $\vert n_\a \vert \ge 1 $,
may then  take negative values
at $ M_X$. When  these values are  large  enough in absolute value,
as for the case of large $\vert n_\a \vert $, the repulsive  contributions of
gaugino  interactions, which also get smaller there because of
the reduced  gauginos masses,
are unable to  flip the  physical scalars masses sign to positive. This
results in drastically reduced averaged traces.
Such values of
$\t $ should be discarded since the associated
vacuum would then lead to tachyonic particles.
The main agent to avoid such negative squared masses are
gaugino loops.  These are more effective
for squarks than for leptons. The  contributions of Yukawa couplings
oppose those of gauginos and are obviously larger for
u-squarks  than  d-squarks  and sleptons. As a result, the upper bounds on
$\cos \t $ from the constraint of the absence of tachyonic particles
are mainly
determined by the sleptons masses. These bounds
depend somewhat on  the \mow assignments.
 For ${\t \over \pi }   $ or $\vert {\t \over \pi} -1 \vert > 0.2 $
no tachyonic particles are present in nearly all three  cases described above.
This is the reason why we have deleted these intervals in plotting the figures.
Finally, we observe that the chargino (neutralino)
minimal mass eigenvalues (figure (f))
lie at $3  \quad (\ud )$ times $m_g$ with a $\t -$dependence
similar to that of the wino mass.

The parameter $m_g$ essentially
fixes the overall scale for most  dimensionful quantities  in spite of
the fact that the ratio ${m_g\over m_Z}$ is  also  relevant  to
the  \ew breaking  condition.
Increasing $m_g$ in the interval  $ [100-1200]$ GeV,
affects slightly $\tan \b $ and negligibly the dimensionless  parameters,
$A^x, B$,
while all superpartner mass parameters essentially scale linearly   with $m_g$.
This is clearly demonstrated by figure 3.

Let us  briefly describe the effect of  changing
the interval of variation of the angle
$\t $ from the first and second quadrants,
corresponding to the choice made above,
to the third and fourth quadrants. This
change leaves scalar masses unchanged but
 modifies the  interference of moduli and dilaton contributions in the other
 parameters, flipping the signs of  $A_x $,  $B$ and leading to
 gaugino masses of negative sign. Naturally
 we need  to flip the signs of gaugino  masses to positive by performing
a phase  redefinition  of the  Majorana spinor
fields of gauginos. The results for the interval of angles $\t \ge \pi $
are not qualitatively different from those above in the interval $\t \le \pi$.

{\bf 3.5 Flavor changing parameters }

The flavor changing parameters are dimensionless  and hence essentially
independent of $m_g$.
Their  variation with Goldstino angle
is depicted in figure 1 for
case I-a and figure 2 for case III-a (figures (g)-(l)).
Let us observe first that because absolute values are plotted here, the sudden
changes of slopes  featured by certain curves are simply due to  changes of
sign in
the  corresponding flavor changing parameters.
The frequent occurrence  of such sign flips, as a function of $\t $,
reflects the presence of large cancellations between the boundary conditions
and the radiative corrections contributions which vary in  opposite
sense with variable  $\t $.

Considering first the unmixed chirality  parameters, we see that
they are  decreasing functions of
$\vert \sin \t \vert$; sleptons
have the fastest slopes  and d-squarks come next.
The fast decrease near small $\t $ or $\pi -\t $
is a combined effect of the decreasing
averaged scalars  masses there   and  the increasing generation dependent
boundary conditions (cf. eqs.(36), (40)).  Due to the specific
structure of the  fermions transformation matrices, the parameters
$(\d_{MM})_{ij}$
are roughly
proportional to  the differences of \mow $ n_i-n_j$.
Turning next to the mixed chirality
parameters, we see that they   vary  rather  slowly with $\t $.
This is explained
by the fact that  the
non-universal contributions  vary in  an opposite sense  from the
gaugino contributions, as previously discussed in subsection
3.2. Sleptons have  the largest  slopes because
of the smaller gauginos contributions there.

A comparison with  experimental bounds in Table 1 reveals that our predicted
values in figures 1 and 2
lie safely below the  experimental ones,
with the exception of $\d^x_{12} $ and $(\d^x_{LR})_{12}$ for $x=d,e$.
We can infer from this comparison lower bounds on $\t $ and $\pi -\t $.
For case I-a, the lower bounds
in the d-squarks case lie  approximately at $0.2$, while the entire interval
of $\t $ angles is excluded in the  sleptons case.
For case III-a, the squarks case entails a larger lower bound  of $0.4$,
while the sleptons case  again excludes all angles.
The excluded intervals in $\t $   increase in proportion to the size of
perturbations  away from universality of modular weights and also to the
size of \mow themselves.
For  the mixed  chirality parameters, the model predictions largely
exceed experimental bounds, except for $(\d^e_{LR})_{12}$
for which predictions  lie an order
of magnitude  above the bounds  for the entire  $\t $ interval.

For a refined comparison  one needs to rescale  experimental
bounds in Table 1 by factors
$ (\tilde m /1000 GeV )$ for squarks
and $ (\tilde m /70 GeV )^p$  for sleptons, with $p=2 $ for $\d^e_{MM}$
and $p=1$ otherwise, using the predicted  average scalar masses given by
figures (d).  This correction induces on the squarks bounds
a reduction by  factors of average
size  $2 $ and for sleptons  an enhancement  by factors  of average size
$10 \quad (3)  $ for unmixed (mixed) chirality parameters. This  will
reduce the predictions for squarks, hence will increase the
lower bounds on $\t $ and $\pi -\t $,  but will partially fill in
the gap between predictions and experimental bounds  for sleptons.
 The rescaling  \wrt the variable
$x$ is not very important for  squarks, for which the model
predictions are compatible with
$x\approx 1$.   For sleptons, however,  a wide gap
separates the values occurring in predictions,
$x_\g \approx 1 $, from that  used   in Table 1, $x_\g \approx 0.02 $.
Since a rescaling of the experimental bounds in Table 1
to $x_\g  =1$  is expected to raise these   bounds
by a factor of order unity [34], this would not modify the above conclusions.
We have not  pursued here a more quantitative study of the $x_\g $
dependence.

It is interesting to determine the maximal perturbations away from
universality  of \mow  allowed by the  condition of absence of tachyonic
particles and
the bounds on flavor changing parameters in the first and second generations.
We find that perturbing  the solutions in ref.[8] for \mow
by  a generation non-universality  exceeding two units results
in  flavor changing parameters larger than
experimental  over the entire interval of $\t $. This  would
essentially rule out the
model. Detailed results for various cases are presented in
Table 2. The three versions of case I are
characterized by degenerate first and second generations.
The one unit differences of modular weights \wrt  the
third generation results in
contributions  of size $10^{-3}$, hence roughly the same order as the radiative
corrections (compare I-a and I-a').
The  comparison of results  between cases I-a and I-a' and II-b and II-b'
shows us that the radiative corrections from first and second
generations  may contribute at about the same  level as the boundary
conditions.
Certain changes of signs reflect the presence of
strong cancellations between radiative corrections
and boundary conditions contributions.
The results for  case II show that mixed chirality  are proportional
to difference of \mow , since going from II-a to II-b enhances the
parameters by  factor $4-10$. The right chirality parameters
are more strongly enhanced
because of stronger cancellations effects present there.
Finally, we note that  the mixed chirality parameters are remarkably stable
within factors $2-3$.

In order to reduce the flavor changing parameters two main possibilities
can be envisaged. One can eventually  assign untwisted string modes, with
$n_\a =-1$, to
all three generations or to a subset.
This choice  might  conflict  with the requirement of obtaining large  scale
hierarchies in the  fermionic mass matrices.  An alternative possibility is
to allow for a non-vanishing goldstino   matter component. This could
significantly reduce the non-universality, since,   as seen from   eq.(10)
the non universal contributions  involve a $\cos^2\t_A $  factor
reduction which is  stronger  than the opposing  gauginos loops
universality  dilution  involving $\cos \t_A$.
In this case one would reduce both squarks and sleptons parameters.

{\bf 4. Conclusions and outlook }

The main issue  adressed in this paper concerned
the \fcnc constraints imposed
on  standard-like orbifold models using the  duality
symmetry approach  initiated in refs.[8,19].
We have obtained detailed
numerical predictions for the \loe spectrum and flavor changing parameters,
essentially covering the  entire parameter space of the
model. Our results complement those of ref.[19] and establish  that all
experimental bounds for masses of superpartners scalars and fermions
can be obeyed provided the gravitino mass is  chosen
to lie above roughly   $  200 $ GeV.
While the flavor changing parameters are
controlled at unification scale by a strong  dependence
on \mow and goldstino angle
the residual  contributions left out  after radiative corrections  are
taken into account
are proportional to \mow  (differences)   and  have a smoothed dependence
on $\t $.
 The  relevant  flavor changing parameters  in the comparison with existing
 experimental bounds are those   associated with the
 first and second generations, especially   sleptons parameters. The
 experimental bounds can be comfortably
 respected  if   $\t $ or $ \pi -\t $  lie  above $0.3$ and  if  non-universal
 perturbations  are chosen for \mow
  not exceeding two units away from  values in solutions
 compatible with anomalies cancellation and  gauge \ccs unification.
Lower bounds of same  size are also required in order to avoid  tachyonic
particles.
The main features of the model which  imply these properties
are the large gauginos and scalars  masses in comparison to the gravitino mass.
The framework in which  the present work   has been developed involves
three main shortcomings which we briefly review here in order to motivate
a future improved   treatment of the problem.

{\it (i) Off-diagonal couplings in the matter K\"ahler potential}.
Terms of form, $\d K= f_{\a \bar \b } (M, \bar M) A_\a A^\dagger _\b +c.c. $,
with $\a , \b $ representing  identical  gauge group \loe modes
but in different generations, are not excluded in general, although
these are  generically absent in  orbifold models. Even when present, such
terms can always be transformed away by a linear transformation of
superfields, which however will  affect the Yukawa couplings. Off-diagonal
matter K\"ahler potentials can also arise  if
massless string modes mix, so that \loe fields are identified
with linear combinations of string modes.  Such a situation is commonly
encountered when one exploits flat directions to assign VEVs to certain
scalars designed to reduce the  rank of the gauge group or to
truncate the massless spectrum.

{\it (ii) Relative alignment of
scalars mass and Yukawa coupling matrices in generation space.}
On side of  gaugino loops radiative corrections,   this is
the next important item  needed to
explain naturally small flavor changing parameters. Although the
structure of the fermions and scalars superpartners matrices  are
sensitive to different characteristics
of orbifolds (\mow versus fixed points) it would be desirable to demonstrate
on specific examples whether some relationship between them
exists and to  demonstrate
its properties. While the threat from a
minimal non-universality which is restricted to
diagonal  elements of mass matrices has been  partially
discarded through our present study, one still needs to worry about
off-diagonal terms.  Such contributions are difficult to avoid when
non-renormalizable interactions are taken into account.
It is  then necessary to verify whether the occurrence of
large  hierarchies in scalars  mass matrices  and in Yukawa
coupling matrices are compatible and if this can be achieved in orbifold
models  without  postulating
an horizontal symmetry.

{\it (iii) CP-violation bounds}. Well-known mechanisms contributing to
imaginary parts of mass shifts in the neutral mesons  systems or to
electric dipole moments provide estimates for the
imaginary  parts of flavor changing
parameters which are an order of magnitude  smaller than  the experimental
bounds for real parts. Approximate calculations of the relevant
CP-violating phases, $\phi_A =Im (A M_a), \quad  \phi_B = Im (B \mu )$,
indicate the need of assigning small values to the complex phases in
the scalar or auxiliary components  of superfields.
If such small contributions are  confirmed by a more quantitative study,
this would then
require that dilaton and moduli fields VEVs should stabilize
very close to real values,  a property which seems to be realized in
a natural way in descriptions of  \susy breaking  incorporating duality
symmetries.

{\bf Acknowledgments}

We are thankful to Carlos Savoy for helpful discussions and suggestions.

\parindent=0 true cm

\def \pr  { Phys. Rev. }
\def \np { Nucl. Phys. }

\def \pl { Phys. Lett. }

{\bf REFERENCES }

1. A. Giveon, M. Porrati and  E.  Rabinovici, Phys. Rep. 244, 77 (1994)

2. S. Ferrara, D. L\"ust, A. Shapere and S. Theisen, \pl B225, 363 (1989);
S. Ferrara, D. L\"ust and S. Theisen, \pl B233,147 (1989)

3.  L. Dixon, V. Kaplunovsky and J. Louis, \np  B329, 27 (1990);

4. M. Cveti\v c and J. Louis and B. A. Ovrut,\pl B206, 227 (1988); ibidem
\pr D40, 684 (1989);
M. Cveti\v c, J. M.  Molera and B. A. Ovrut, \pr D40, 1146 (1989);
J. M. Molera and B. A. Ovrut, \pr D42, 2683 (1990); D. Bailin,
A. Love and W. A. Sabra, \pl B275, 55 (1992)

5. M. Dine, P. Huet and N. Seiberg, \np B322, 301 (1989)

6. L. E. Ib\'a\~nez, W. Lerche,
D. L\"ust and S. Theisen,   \np B352, 435 (1990)

7. J. Lauer, J. Mas and H. P. Nilles, \pl B226, 251 (1989);
\np B351, 353 (1991); E. J. Chun, J. Mas, J. Lauer
and H. P. Nilles, \pl B233, 141 (1989); W. Lerche, D. L\"ust and N. P. Warner,
\pl 231, 417 (1989)

8.  L.E. Ib\'a\~nez and D. L\"ust, \np B382, 305 (1992)

9.  L. Dixon, V. Kaplunovsky and J. Louis, \np  B355, 649 (1991)

10. I. Antoniadis, K. S. Narain and T. R. Taylor, \pl B267, 37 (1989);
I. Antoniadis, E. Gava and  K. S. Narain, \np B383, 93 (1992);
I. Antoniadis, E. Gava, K. S. Narain and T. R. Taylor, \np B

11. G. L. Cardoso and  B. Ovrut, \np B369, 351 (1992);ibidem, \np B418,
535 (1994); V. Kaplunovsky and J. Louis, \np B422, 57 (1994)

12.  J.-P. Derendinger, S. Ferrara,  C. Kounnas, F.
Zwirner, \pl B271, 307 (1991); ibidem B372, 145 (1992)

13. S. Ferrara, N. Magnoli, T.R. Taylor and G. Veneziano, \pl B245, 409 (1990);
A. Font, L. E. Ib\'a\~nez, D. L\"ust and F. Quevedo, \pl B245, 401 (1990);
T. R. Taylor and G. Veneziano, \pl B212, 147 (1988); V. Krasnikov, \pl B193, 37
(1987)

14. M.  Cveti\v c, A.  Font, L.E. Ib\'a\~nez, D. L\"ust  and F. Quevedo,
\np B361, 194 (1991)

15. H. P. Nilles and M. Olechowsky, \pl B248, 268 (1990); P. Bin\'etruy and
M. K. Gaillard, \pl B253, 119 (1991); J. A. Casas, Z.  Lalak, C.
Mu\~noz and G. G. Ross, \np  B347, 243 (1990);
D. L\"ust and T. R. Taylor, \pl 253, 335 (1991);
S. Kalara, J. Lopez and D. Nanopoulos, \pl B275, 304 (1992)

16. V. Kaplunovsky and J. Louis, \pl  B306, 269 (1993);  R.
Barbieri,  J. Louis and M. Moretti, \pl B312, 451 (1993)

17. B. de Carlos, J.A. Casas and C. Mu\~noz, \pl B299, 234 (1993)

18. B. de Carlos, J.A. Casas and C. Mu\~noz, \np B399, 623 (1993)

19. A. Brignole, L.E. Ib\'a\~nez, C. Mu\~noz, \np B42, 125 (1994)

20. F. Gabbiani and A. Masiero, \np B322, 235 (1989)

21. S. Bertolini, F.  Borzumati, A. Masiero and G. Ridolfi, \np
B353, 591 (1991)

22. K.  Inoue et al., Prog. Theor. Phys. 68, 927 (1982);
L. Alvarez-Gaum\'e, J. Polchinski and M. Wise, \np B221,
495 (1983); L. E. Ib\'a\~nez, \np B218, 524 (1983);
L. E. Ib\'a\~nez and C. Lopez, \pl B126, 54 (1983); ibid. \np B233, 511 (1984)
J.-P. Derendinger and C. A.  Savoy, \np B237, 307 (1984);
A. Bouquet, J. Kaplan and C. A. Savoy, \pl B148, 69 (1984)

23. L. E. Ib\'a\~nez and C. Lopez, \np B233, 511 (1984)

24. H. Georgi, \pl B169, 231 (1986)

25.  L. Hall, A.  Kostelecky and S. Raby, \np  B267, 415 (1986)

26. M. Dine, A. Kagan and S. Samuel, \pl B243, 250 (1990)

27. Y. Nir and N. Seiberg, \pl B309, 337 (1993);
P. Pouliot and N. Seiberg, \pl  B318, 169 (1993); M. Leurer, Y. Nir
and N. Seiberg, \np B420. 468 (1994)

28. M. Dine, A. Kagan and  R. Leigh, \pr  D48, 4269 (1993)

29. D.  B. Kaplan  and M. Schmaltz, \pr D49,3741 (1994)

30. L. E. Ib\'a\~nez and G. G. Ross, \pl B332, 100 (1994)

31. D. Matalliotakis and H.P. Nilles, TUM-HEP-201/94;
J. Kim and H. P. Nilles, MPI-Pht/94-39

32. D. Choudhury, F. Eberlein, A. Konig, J. Louis and  S. Pokorski, preprint
MPI-PhT/94-51, LMU-TPW 94-12

33. A. Lleyda and C. Mu\~noz, \pl B317, 82 (1993)

34. J. S. Hagelin, S. Kelley and T. Tanaka, \np B415, 293 (1994)

35. A. Font, L. E. Ib\'a\~nez, F. Quevedo and A. Sierra, \np B331, 421 (1990);
L. Dixon, Superstrings, unified theories and cosmology, Trieste 1987
Summer Workshop, eds. G. Furlan et al., (World Scientific, 1988)

36. J. A. Casas, F. Gomez and C. Mu\~noz, Int. J. Mod. Phys. A8, 455 (1993);
P. Mayr and  S. Stieberger, \np B407, 725 91993); D. Bailin,
A. Love, W. A. Sabra and S. Thomas, \pl B320, 21 (1994)

37. L. E. Ib\'a\~nez and D. L\"ust, \pl B267, 51 (1991)

38. L. Dixon, J. A. Harvey, C. Vafa and E. Witten, \np B261, 651 (1985);
ibidem \np B274, 285 (1986)

39. J. E.  Casas, F. Gomez  and C. Mu\~noz, \pl B292, 42 (1992)

40. E. Cremmer, S. Ferrara, L. Girardello and A. Van Proeyen,
\np B212, 413 (1989);

41. L. Dixon, D. Friedan, E. Martinec and S. Shenker, \np B271, 189 (1986);
S. Hamidi and C. Vafa, \np B279, 465 (1987); T. T. Burwick, R. K. Kaiser
and H. F. Muller, \np B   , 689 (1991); D. Bailin,
A. Love and W. A. Sabra,
\np B416, 539 (1994)

42. L. E. Ib\'a\~nez and G.G. Ross, Perspectives in Higgs
Physics, ed. G. Kane (World Scientific)

43. J.A. Casas and C. Mu\~noz, \np B332, 189 (1990)

44.  P. Ramond,  R.G. Roberts  and G.G.  Ross, \np  B406, 19 (1993)

45. L. E. Ib\'a\~nez, D. L\"ust  and G.G. Ross, \pl B272, 251 (1991)

\vfill\eject
{\bf Tables Captions}

{\bf Table 1.} Experimental upper bounds for  the $\d^x_{MN}$ matrices of
up-squarks (1st line), d-squarks (2d-4th lines) and sleptons (5th line).
Our  sources are  the works of  Gabbiani and
Masiero (GM) [20], Nir and Seiberg (NS) [27] and
Hagelin et al., (HKT) [34]. We set the average squarks
mass to $\tilde m=1000 $ GeV and sleptons mass to  $\tilde m =70 $ GeV.
The parameters   $x_3= ({M_3\over \tilde m})^2 $ (squarks)
 or $ x_\g =({M_{\tilde \g }\over \tilde m})^2$ (sleptons) are  set to
 $ 1$ in the bounds of NS [27] and HKT [34] and to $ x_3=0.49,
 x_\g \approx 0.029 $ in those of GM [20].
(Contributions
arising from $A $ and $\mu $ terms  in ref.[20] are neglected.)
The notation $\d_{ij} $ is reserved to bounds obtained from
box diagrams contributions proportional to the products $\d_{LL} \d_{RR} $.
Likewise, bounds on $\d_{LR}$ are obtained from contributions
involving the products $[(\d_{LR} )_{ij} (\d_{LR})_{ji}]^\ud$.
For neutral $K\bar K , B\bar B , D \bar D $ systems,  the bounds  on
$(\d_{MN})_{ij} $  scale as $\tilde m$ at fixed $x$. For  decays
$q_i \to q_j + \g ,
l_i \to l_j + \g $,  the bounds for $(\d_{MM})_{ij}$  scale as $\tilde m^2 $
and  those for $(\d_{LR})_{ij}$ as $\tilde m$, at fixed $x$.
Thus, except for the  sleptons bounds $\d_{MM}$ here, which scale as $ \tilde
m^2$,
all other  bounds  quoted in the table  scale as $ \tilde m$.
Note that the  new  Cleo  measurements  for $b\to s +\g $
yield the bound $(\d^d_{LR})_{23} =0.028$.

{\bf Table 2.} Flavor changing parameters for d-squarks (top half) and sleptons
(bottom half) for  fixed $m_g =200 GeV, \t /\pi = 0.35$.
The  prime indicates that  we have
omitted radiative corrections in first and second generations which
are described  in the leading logarithm approximation of eq.(23).
The  various cases
described in Section 3.3 are identified in the first column.
The \mow in the order $\a =[Q;U^c;D^c;L;E^c]$ are: case I-a:
$n_\a =-[221;332;221;223;223]; $ case II-a: $n_\a=-[111;133;111;333;133]$;
caseII-b: $n_\a =-[023;133;023;023;133];$ case III-a:$n_\a =-[123]$;
case III-b:$n_\a =-[112]$;
\vskip 1cm
{\bf Figures  Captions}

{\bf Figure  1.}  Plot as a function of $\t /\pi $ for $ m_g=200 $ GeV
of the model parameters in Case I-a: (a) $\tan \b $; (b) $\mu (0)$;
(c)  \sm gauginos masses $M_a(t_Z)$  in the order
$ a=3,2,1$ from top to bottom ; (d) Average trace of left times right chirality
mass submatrices, $ \tilde m^x = [ Trace(\tilde M^2_{LL}(t_Z))
Trace(\tilde M^2_{RR}(t_Z)) ]^{1\over 4},$ for d-squarks (continuous),
u-squarks (dot-dashed)  and  sleptons (dashed);
(e) \susy breaking parameters $A_x (t_Z) $ for  x=t (continuous),
b (dot-dashed), $\tau  $(dashed), and  $B(t_Z) $ (dotted);
(f) minimum  masses  of charginos (continuous) and neutralinos
(dot-dashed); (g)-(i) $ \d^x_{ij}=\vert
(\d^x_{LL})_{ij} (\d^x_{RR})_{ij})\vert ^\ud , \quad [x=d,e,u]  $  for
 $ (i,j)=(1,2) $ (figure (g)), $(1,3) $(figure (h))  and $ (2,3) $
 (figure (i)); (j)-(l) $\d^{'x}_{ij}=\vert
(\d^x_{LR})_{ij} (\d^x_{LR})_{ji})\vert ^\ud ,   \quad [x=d,e,u] $ for
 $ (i,j)=(1,2) $ (figure (j)), $(1,3) $(figure (k)) and $ (2,3) $ (figure (l)).
Curves in   each of figures (g)-(l) refer to  d-squarks (continuous),
sleptons (dot-dashed) and u-squarks (dashed).
Note that in figures (g)-(l) the discontinuities in slopes occurring for
certain curves signal the occurrence in these semi-logarithmic
plots  of changes of signs for quantities in ordinate.

{\bf Figure  2.} Same conventions  as  figure 1 in  Case III-a.

{\bf Figure  3.} Plot versus $ m_g $ for  fixed $\t =0.35 \pi$ of \loe
parameters
in Case I-a, using  the same conventions
for curves as in figure 1.

\vfill\eject
\def \d {\delta }
\centerline {\bf TABLE 1 }
\vskip 0.5cm
$$\vbox {\settabs  10\columns
\+ Sector & $(\d_{MM})_{12}$&$(\d_{MM})_{13}$&$(\d_{MM})_{23}$
&$\d_{12}$& $\d_{13}$ & $\d_{23}$
& $ (\d_{LR})_{12}$  & $(\d_{LR})_{13}$ & $(\d_{LR})_{23} $ \cr
\+  \cr
\hrule
\+  \cr
\+ U (NS) & $0.1 $  &&& $0.04$ &&& $0.06$ &&\cr
\+ ref.[27]\cr
\+  \cr
\hrule
\+ \cr
\+  D (NS) & $0.05 $  & $0.1 $ && $0.006$ & $ 0.04 $&& $0.008$ & $ 0.06 $ &
$0.04$ \cr
\+ ref.[27]\cr
\+  \cr
\+  D (GM) & $0.035 $  & $0. 23$ && $0.0051$ & $ 0.074 $&& $0.007$ & $ 0.1 $ &
$0.12$ \cr
\+ ref.[20]\cr
\+  \cr
\+  D (HKT)& $0.1 $  & $0.27$ & $ 47 $ & $0.006$ & $ 0.073 $&& $0.0044$
& $ 0.14 $ & $0.071$ \cr
\+ ref.[34]\cr
\+  \cr
\hrule
\+  \cr
\+ L (GM)& $8.9610^{-4} $ & $7.67 $& $7.08$& &&& $1.5510^{-6}$&
$2.2410^{-1}$& $2.0610^{-2}$  \cr
\+ ref.[20]\cr
\+  \cr
\hrule
}$$
\vfill\break
\centerline {\bf TABLE 2 }
\vskip 0.5cm
$$\vbox {\settabs  10\columns
\hrule
\+ \cr
\+Case & $(\d_{LL})_{12}$&$(\d_{LL})_{13}$&$(\d_{LL})_{23}$
&$(\d_{RR})_{12}$& $(\d_{RR})_{13}$ & $(\d_{RR})_{23}$
& $ (\d_{LR})_{12}$  & $(\d_{LR})_{13}$ & $(\d_{LR})_{23} $ \cr
\+ \cr
\hrule
\+ \cr
\+ {\bf D-SQUARKS } \cr
\+ Ia& $3.210^-{}^3 $ & $-3.6 10^-{}^4$& $3.0 10^-{}^3$& $3.7 10^-{}^4$
&$4.710^-{}^4$&$-6.610^-{}^4$&
$9.2 10^-{}^5$& $9.4 10^-{}^5 $ & $4.2 10^-{}^4$ \cr
\+ Ia'& $-4.010^-{}^4 $ & $-5.1 10^-{}^4$& $3.7 10^-{}^3$& $-4.3 10^-{}^4$
&$4.310^-{}^4$&$-5.110^-{}^4$&
$9.2 10^-{}^5$& $9.4 10^-{}^5 $ & $4.2 10^-{}^4$ \cr
\+ Ib& $3.210^-{}^3 $ & $-8.5 10^-{}^4$& $5.2 10^-{}^3$& $3.3 10^-{}^5$
&$4.110^-{}^4$&$-4.210^-{}^4$&
$7.9 10^-{}^5$& $8.4 10^-{}^5 $ & $3.7 10^-{}^4$ \cr
\+ \cr
\hrule
\+ \cr
\+ IIa& $2.910^-{}^3 $ & $-6.4 10^-{}^4$& $4.3 10^-{}^3$& $6.7 10^-{}^5$
&$3.110^-{}^4$&$6.810^-{}^5$&
$9.8 10^-{}^5$& $1.1 10^-{}^4 $ & $4.7 10^-{}^4$ \cr
\+ IIb& $1.110^-{}^2 $ & $-9.7 10^-{}^4$& $-5.8 10^-{}^3$& $8.4 10^-{}^3$
&$1.310^-{}^4$&$8.810^-{}^4$&
$1.2 10^-{}^4$& $1.2 10^-{}^4 $ & $5.2 10^-{}^4$ \cr
\+ IIb'& $7.310^-{}^3 $ & $-1.1 10^-{}^3$& $6.6 10^-{}^3$& $7.8 10^-{}^3$
&$1.110^-{}^4$&$9.710^-{}^4$&
$1.2 10^-{}^4$& $1.2 10^-{}^4 $ & $5.2 10^-{}^4$ \cr
\+ \cr
\hrule
\+ \cr
\+ IIIa& $5.810^-{}^3 $ & $-1.3 10^-{}^3$& $7.2 10^-{}^3$& $3.0 10^-{}^3$
&$1.610^-{}^4$&$7.210^-{}^4$&
$9.2 10^-{}^5$& $1.0 10^-{}^4 $ & $4.5 10^-{}^4$ \cr
\+ IIIb& $1.910^-{}^3 $ & $-1.2 10^-{}^3$& $6.8 10^-{}^3$& $-2.3 10^-{}^4$
&$1.610^-{}^4$&$7.210^-{}^4$&
$8.5 10^-{}^5$& $1.0 10^-{}^4 $ & $4.6 10^-{}^4$ \cr
\+  \cr
\hrule
\+  \cr
\+ {\bf SLEPTONS } \cr
\+  \cr
\+  Ia& $7.110^-{}^3 $ & $2.8 10^-{}^4$& $-7.7 10^-{}^4$& $-1.7 10^-{}^4$
&$-4.810^-{}^4$&$1.110^-{}^2$&
$3.4 10^-{}^5$& $4.1 10^-{}^5 $ & $6.0 10^-{}^4$ \cr
\+  Ia'& $2.310^-{}^4 $ & $-1.510^-{}^4$& $5.8 10^-{}^3$& $-2.2 10^-{}^4$
&$-4.810^-{}^4$&$1.110^-{}^2$&
$3.4 10^-{}^5$& $4.1 10^-{}^5 $ & $6.0 10^-{}^4$ \cr
\+  Ib & $1.010^-{}^2 $ & $4.110^-{}^4$& $-2.8 10^-{}^3$& $-1.8 10^-{}^4$
&$-6.910^-{}^4$&$1.410^-{}^2$&
$3.6 10^-{}^5$& $4.3 10^-{}^5 $ & $6.3 10^-{}^4$ \cr
\+  \cr
\hrule
\+ \cr
\+ IIa& $7.010^-{}^3 $ & $6.8 10^-{}^4$& $-6.1 10^-{}^3$& $2.6 10^-{}^2$
&$3.110^-{}^4$&$3.710^-{}^5$&
$4.7 10^-{}^5$& $3.2 10^-{}^5 $ & $4.5 10^-{}^4$ \cr
\+ IIb& $2.210^-{}^2 $ & $3.4 10^-{}^4$& $-1.8 10^-{}^4$& $2.6 10^-{}^2$
&$3.110^-{}^4$&$6.110^-{}^5$&
$5.5 10^-{}^5$& $3.3 10^-{}^5 $ & $4.9 10^-{}^4$ \cr
\+ IIb'& $1.410^-{}^2 $ & $-1.3 10^-{}^4$& $6.0 10^-{}^3$& $2.6 10^-{}^2$
&$3.010^-{}^4$&$1.110^-{}^4$&
$5.5 10^-{}^5$& $3.3 10^-{}^5 $ & $4.9 10^-{}^4$ \cr
\+  \cr
\hrule
\+ \cr
\+ IIIa& $1.810^-{}^2 $ & $3.5 10^-{}^4$& $4.7 10^-{}^4$& $1.8 10^-{}^2$
&$-7.510^-{}^4$&$1.510^-{}^2$&
$4.8 10^-{}^5$& $4.0 10^-{}^5 $ & $6.0 10^-{}^4$ \cr
\+ IIIb& $6.910^-{}^3 $ & $2.1 10^-{}^4$& $2.2 10^-{}^3$& $-2.0 10^-{}^4$
&$-6.410^-{}^4$&$1.310^-{}^2$&
$3.1 10^-{}^5$& $3.5 10^-{}^5 $ & $5.4 10^-{}^4$ \cr
\+ \cr
\hrule
}$$

\end